\documentclass[notitlepage]{article}

\usepackage{authblk}
\usepackage{graphicx}
\usepackage{epstopdf}
\usepackage{subcaption}
\usepackage{amsmath}
\usepackage{multirow}
\usepackage{booktabs}
\usepackage{topcapt}
\usepackage{xcolor}
\usepackage{amssymb}
\usepackage{times}
\usepackage{comment} 

\usepackage{titling}

\newcommand{\specialcell}[2][c]{%
  \begin{tabular}[#1]{@{}c@{}}#2\end{tabular}}

\newcommand{\nop}[1]{}

\usepackage{color}

\begin{document}
\title{Rating Effects on Social News Posts and Comments}
\author{Maria Glenski}
\author{Tim Weninger}
\affil{Department of Computer Science and Engineering\\University of Notre Dame}
\date{}                     
\setcounter{Maxaffil}{0}
\renewcommand\Affilfont{\itshape\small}
    \maketitle
    \begin{abstract}
At a time when information seekers first turn to digital sources for news and opinion, it is critical that we understand the role that social media plays in human behavior. This is especially true when information consumers also act as information producers and editors through their online activity. In order to better understand the effects that editorial ratings have on online human behavior, we report the results of a two large-scale {\em in-vivo} experiments in social media. We find that small, random rating manipulations on social media posts and comments created significant changes in downstream ratings resulting in significantly different final outcomes. We found positive herding effects for positive treatments on posts, increasing the final rating by 11.02\% on average, but not for positive treatments on comments. Contrary to the results of related work, we found negative herding effects for negative treatments on posts and comments, decreasing the final ratings on average, of posts by 5.15\%  and of comments by 37.4\%. Compared to the control group, the probability of reaching a high rating ($\ge$2000) for posts is increased by 24.6\% when posts receive the positive treatment and for comments is decreased by 46.6\% when comments receive the negative treatment.
\end{abstract}

\section{Introduction}

We often rely on online reviews contributed by anonymous users as an important source of information to make decisions about which products to buy, movies to watch, news to read, or even political candidates to support. These online reviews replace traditional word-of-mouth communication about an object's or idea's quality~\cite{Chevalier2006}. The sheer volume of new information being produced and consumed only increases the reliance that individuals place on anonymous others to curate and sort massive amounts of information. Because of the economic and intrinsic value involved, it is important to consider whether this new mode of social communication can successfully harness the wisdom of crowd to accurately aggregate individual information. 

What is becoming known as \textbf{collective intelligence} bares the potential to enhance human capability and accomplish what is impossible individually~\cite{Woolley2010,Brabham2008}. For example, more than a century ago the experiments of Francis Galton determined that the median estimate of a group can be more accurate than estimates of experts~\cite{Galton1907}. Surowiecki's book \textit{The Wisdom of the Crowds} finds similar examples in stock markets, political elections, quiz shows and a variety of other fields where large groups of people behave intelligently and perform better than an elite few~\cite{Surowiecki2005}. However, other experiments have shown that when individuals' perceptions of quality and value follow the behavior of a group, the resulting \textbf{herd mentality} can be suboptimal for both the individual and the group~\cite{Bikhchandani1992,Hirshleifer1995,Lorenz2011}.

By relying on the judgements of others, we may be susceptible to malicious ratings with some ulterior motive. Unfortunately, there is a gap in our knowledge and capabilities in this area, including untested and contradictory social theories. Fortunately, these gaps can be filled using new experimental methodologies on large, socio-digital data sets. The main idea is to determine if these socio-digital platforms produce useful, unbiased, aggregate outcomes, or (more likely) if, and how, opinion and behavior is influenced and manipulated. Work of our own and recent tangential experiments~\cite{Weninger2015,Muchnik2013,Lakkaraju2013,Salganik2006} suggest that decisions and opinions can be significantly influenced by minor manipulations, yielding different social behavior.

\vspace{.3cm}
The main focus of the present work is the determine what effect, if any, does malicious voting behavior have on social news posts and comments.
\vspace{.3cm}

Unfortunately, causal determinations are difficult to assess. In a closely related experiment, Wu and Huberman measured rating behavior in two different online platforms. The first allowed users to see prior ratings before they voted and the other platform hid the prior ratings until after the user voted. They found that when no information about previous ratings or page views are available, the ratings and user-opinions expressed tend to follow regular patterns. However, in cases where the previous ratings were made known, the user-opinions tended to be either neutral or form a polarized consensus. In the latter case, new opinions tend to reinforce previous opinions and thus become more extreme~\cite{Wu2008}.

\begin{figure*}[t]
    \centering
    \begin{minipage}[b]{\textwidth}
    
        \includegraphics[width=\textwidth]{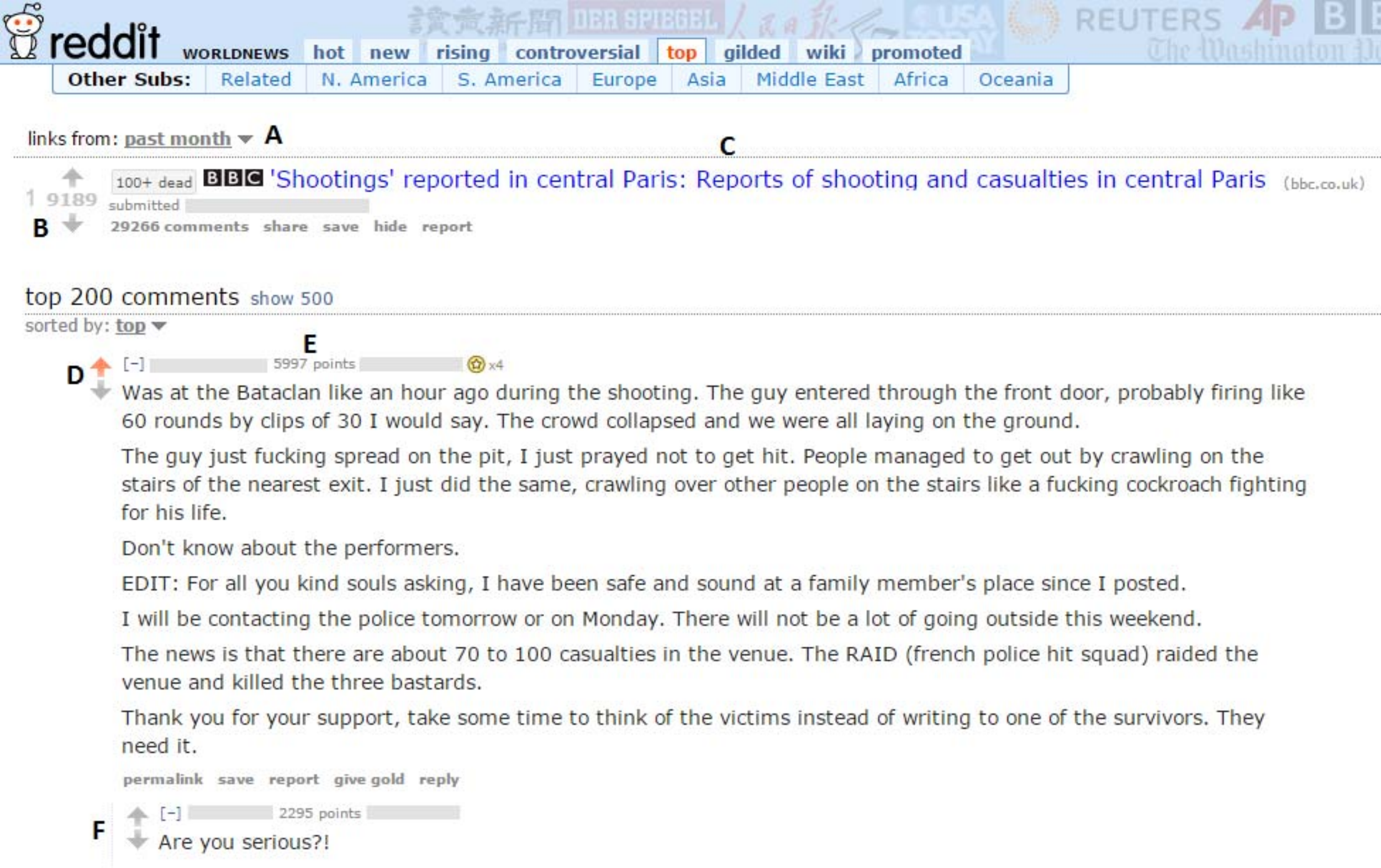}
    \caption{Composite, redacted screenshot of Reddit. (A) There are many possible ranking systems on Reddit; in this image shows the first post when ordered by the ``top''--scored posts within the past month. (B) Authenticated users may up-vote, or down-vote once on any post; the score of a post is congruent to the total number of up-votes minus the total number of down-votes. (C) Each post displays its rank on the far left corresponding to its position in the selected ranking system, a title text, and the host of the linked content on the far right; this post also has 29,266 comments. (D) The top 200 comments are displayed in order as well corresponding to the chosen ranking system and number of points the comment has received; an orangered arrow indicates that the current user up-voted this comment. (E) This comment has a score of 5,997, which is congruent to the number of up-votes minus the number of down-votes that the comment has received. (F) Comment threads are hierarchical such that each comment can have have children, siblings, etc. thus comment orderings are based on their vote score relative to the sibling comments in the thread hierarchy.}
    \label{fig:redditpage}
    \end{minipage}
\end{figure*}

Because of the information overload caused by billions of daily shares, tweets, posts and comments, nearly all social media Web sites have sophisticated ranking algorithms that attempt to identify the relatively few items that their users will find interesting. When new or better items are shared or posted, the ranking systems rely significantly upon user ratings to accelerate the discovery of new or interesting content. Information that is rated positively will be ranked higher, and will therefore be more visible to other users, which further increases the likelihood that it will receive further ratings~\cite{Buscher2009,Payne2014}. 

A recent experiment by Lerman and Hogg further studied the effects that presentation order has on the choices that users make. In this study, several users were asked to read and rate social media posts ranked by various ordering algorithms. Lerman and Hogg found that different ranking systems result in very different outcomes. Random orderings result in the most unbiased ratings, but may show a lot of uninteresting content resulting in poor user engagement. The ``popularity ranking,'' which rated posts by how many previous positive-votes it received led to highly inconsistent outcomes and showed that small early differences in ratings led to inconsistent rating scores~\cite{Lerman2014}.

Social news Web sites represent a stark departure from traditional media outlets in which a news organization, {\em i.e.}, a handful of television, radio or newspaper producers, sets the topics and directs the narrative. Socio-digital communities increasingly set the news agenda, cultural trends, and popular narrative of the day~\cite{Leskovec2009a}. News agencies frequently have segments on ``what's trending,'' entire television shows are devoted to covering happenings on social media, and even live presidential debates select their topics based on popular questions posed to social media Web sites. As this trend continues and grows, the number of blogs, news outlets, and other sources of user generated content has outpaced the rate at which Web users can consume information. Social news Web sites are able to automatically curate, rank and provide commentary on the top content of the day by harnessing the wisdom of the crowd.

The recent popularity of social networks has led to the study of socio-digital influence and popularity cascades where models can be developed based on the adoption rate of friends ({\em e.g.}, shares, retweets). Bakshy {\em et al.}, find that friendship plays a significant role in the sharing of content~\cite{Bakshy2009}. Similarly, Leskovec {\em et al.} were able to formulate a generative model that predicts the size and shape of information cascades in online social networks~\cite{Leskovec2006}. 

However, social media users seem to be unaware of the effects of social manipulation. A recent survey of Reddit users aimed to determine what the sampled community thought drove Reddit-users to up-vote or down-vote various posts. The surveyors expected that the leading indicators would be that users are more likely to up-vote or like 1) content that others have liked, indicating social influence; 2) content that was submitted or contributed by a well known user, indicating trust or model-based bias; or 3) content that is relevant to the user's interests. Contrary to our scientific understanding of social influence, the surveyed users indicated that social influence had little effect on their voting likelihood~\cite{Priestley2015}. Other than the need to raise awareness of the impact of social influence within social media communities, these results suggest that online social media aggregators are a viable testbed for theories of trust and social influence.

Like social networks, online {\em social news} platforms allow individuals to contribute to the wisdom of the crowd in new ways. These platforms are typically Web sites that contain very simple mechanics. In general, there are 4 operations that are shared among social news sites: 

\begin{enumerate}
	\item individuals generate content or submit links to content,
	\item submissions are rated and ranked according to their rating scores,
	\item individuals can comment on the submitted content,
	\item comments are rated and ranked according to their rating scores.
\end{enumerate}
Simply put, social news platforms allow individuals to submit content and vote on the content they like or dislike.

The voting mechanism found in socio-digital platforms provides a type of Web-democracy that is open to everyone. Given the widespread use and perceived value of these voting systems~\cite{Gilbert2013}, it is important to consider whether they can successfully harness the wisdom of the crowd to accurately aggregate individual information.
In our study, we determine what effect, if any, ranking and vote score has on rating behavior. This is accomplished via an {\em in vivo} experiment on the social media Web site, Reddit, by inserting random votes into the live rating system.

Reddit is a social news Web site where registered users can submit content, such as direct posts or links.  Registered users can then up-vote submissions or down-vote submissions to organize the posts and determine the post's position on the site; posts with a high vote score ({\em i.e.}, up-votes -- down-votes) are ranked more highly than posts with a low vote score. Reddit is organized into many thousands of ``subreddits,'' according to topic or area of interest, {\em e.g.}, \texttt{news}, \texttt{science}, \texttt{compsci}, \texttt{datamining}, and \texttt{theoryofreddit}, and posts must be submitted to a subreddit. A user that subscribes to a particular subreddit will see highly ranked posts from that subreddit on their frontpage, which Reddit describes as `the front page of the Internet' and is unique for each user. Figure~\ref{fig:redditpage} illustrates an example post and a small piece of its comment section.

As in most social media Web sites, users are free to comment on the posts. Reddit has a unique commenting framework that ranks comments based on their scores relative to their sibling comments. For instance, all root comments, {\em i.e.}, comments with no parent, are ranked together, and all of the children-comments of some single parent-comment are ranked together. It is possible, even frequent, that a comment deep within the comment thread-tree will have a higher score than its parent or ancestor-comments~\cite{Weninger2013}.

By default, Reddit only displays the top 200 comments, even though it is common for popular posts to receive thousands of comments. Therefore, many comments in popular threads are never viewed, which likely exacerbates the rich-get-richer effect that is already seen in certain ranking systems.

It is important to note that, unlike other online social spaces, Reddit is not a social {\em network}. the notion of friendship and friend-links, like on Facebook, is mostly absent on Reddit. Although usernames are associated with posts and comments, the true identity of registered users is generally unknown and in many cases fiercely guarded. 

In fact, we attempted to find friendship by looking at user-pairs that frequently reply to each other in comments; unfortunately, more than 99.9\% of the comments were in reply to a user that they had never previously replied to. Thus, we typically refer to Reddit a social non-network, and the vast amount of previous social {\em network} literature does not apply.

In the present work, we report the results of two large in-vivo experiments on Reddit; the first ($N=93,019$) up-voted or down-voted posts at random and the second ($N=128,316$) up-voted or down-voted comments at random. Based on these experimental treatments we observe the effects that votes have on the final score of a post or comment as a proxy for observing herding effects in social news. Unlike the experimental study performed by Muchnik \textit{et al.}, and other behavioral studies our experiments: 1) manipulate votes of posts and comments rather than just comments, 2) leverages Reddit's dynamic, score-based ranking system rather than a time-only ranking system, 3) does not involve friendship or the use of social networks, and 4) randomly delays the vote treatment rather than always performing the treatment immediately upon creation.

These differences are significant in that this is the first ever vote manipulation experiment on a global scale, live, working system. The use of randomized trials eliminates concerns about various confounding factors, and we have made our data and analysis scripts available to the community for replication and further research.

\section{Methods}

\subsection{Post Experiment}
During the 6 months between September 1, 2013 and January 31, 2014 a computer program was executed every 2 minutes that collected post data from Reddit through an automated two-step process. First, the most recent post on Reddit was identified and assigned to one of three treatment groups: up-treated, down-treated, or control. Up-treated posts were artificially given an up-vote (a +1 rating) and down-treated posts were given a down-vote (a -1 rating). Up-treatment, down-treatment and the control have an equal likelihood of being selected. Vote treated posts are assigned a random delay ranging from no delay up to an hour delay in intervals of 0, .5, 1, 5, 10, 30 and 60 minutes. Second, each post was re-sampled 4 days later and final vote totals were recorded. 

These treatments created a small, random manipulation signalling positive or negative judgement that is perceived by other voters as having the same expected quality as all other votes thereby enabling estimates of the effects of a single vote while holding all other factors constant. This data collection resulted in 93,019 sampled posts, of which 30,998 were up-treated and 30,796 were down-treated; each treatment type was randomly assigned a delay interval with equal likelihood. Treatments were removed from the vote scores before data analysis was performed, {\em i.e.}, up-treated post-scores were decremented by 1 and down-treated post-scores were incremented by 1.

During the experimental time period, Reddit reported that their up-vote and down-vote totals were ``fuzzed'' as an anti-spam measure; fortunately, they certified that a post's score ({\em i.e.}, up-votes minus down-votes) was always accurate. In July of 2014, after the data gathering phase of this experiment had ended, Reddit removed the vote totals from their Web site and replaced it with a semi-accurate points system; Reddit administrators currently assert that the rankings are always accurate, even though their reported scores may not be.

\subsection{Comment Experiment}
During the 6 months between September 1, 2013 and January 31, 2014 a computer program, separate from the post experiment, was executed every 2 minutes that collected comment data from Reddit through an automated two-step process. First, the most recent comment on the top ranked post ordered by the "rising" ranking algorithm on the Reddit frontpage was identified and assigned to one of three treatment groups: up-treated, down-treated, or control. Up-treated comments were artificially given an up-vote (a +1 rating) and down-treated comments were given a down-vote (a -1 rating). Up-treatment, down-treatment and the control have an equal likelihood of being selected. Vote treated comments are assigned a random delay ranging from no delay up to an hour delay in intervals of 0, .5, 1, 5, 10, 30 and 60 minutes. Second, each comment was re-sampled 4 days later and final vote totals were recorded. 

These treatments produced a score manipulation similar to that of the post experiment, wherein all other factors were held constant enabling a clear causal signal to be measured. This data collection resulted in 96,486 sampled comments, of which 35,704 were up-treated and 31,830 were down-treated; each treatment type was randomly assigned a delay interval with equal likelihood. Treatments were removed from the vote scores before data analysis was performed, {\em i.e.}, final up-treated comment-scores were decremented by 1 and final down-treated comment-scores were incremented by 1.

To our knowledge, comment scores were not fuzzed in the same way that post scores are fuzzed, so absolutely point scores reported here should be accurate.

The voting agents used here were periodically checked to ensure that they had not been blocked or their votes ignored. The voting agent did not target any one type of content or subreddit or content provider, which are among the most common types of vote-spam, therefore, we are certain that all of our votes were counted.

\begin{table}[]
\begin{center}
\begin{tabular}{ccccc}
\hline
Experiment & Up-Treatment & Down-Treatment & Control & Total \\ \hline 
Post & 30,998 & 30,796 & 31,225 & 93,019 \\
Comment & 35,704 &	31,830 &   28,952 & 96,486 	\\
\hline
\end{tabular}
\end{center}
\label{tab:data_summary_table}
\caption{Summary of sample count by treatment for the data collected from September 1, 2013 to January 31, 2014
through the Post and Comment Experiments.}
\end{table}

\section{Results}
We first compared the final vote totals of each treatment group. These findings measure the overall effect that up-treatments and down-treatments have on the overall life of a post or comment.

\begin{figure}[t]  
    \centering
    \begin{minipage}[b]{0.35\textwidth}
        \includegraphics[width=\textwidth]{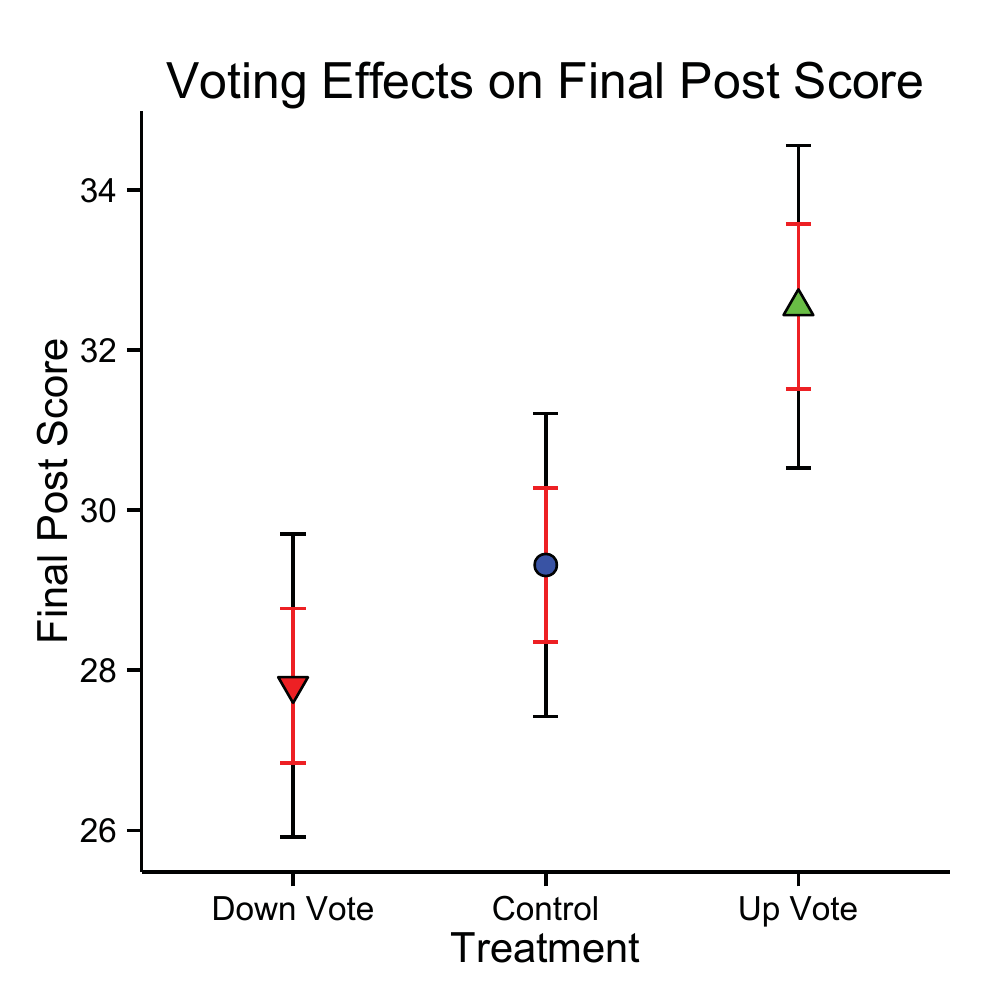}
        \subcaption{Posts}
    \end{minipage}
    ~
    \begin{minipage}[b]{0.35\textwidth}
        \includegraphics[width=\textwidth]{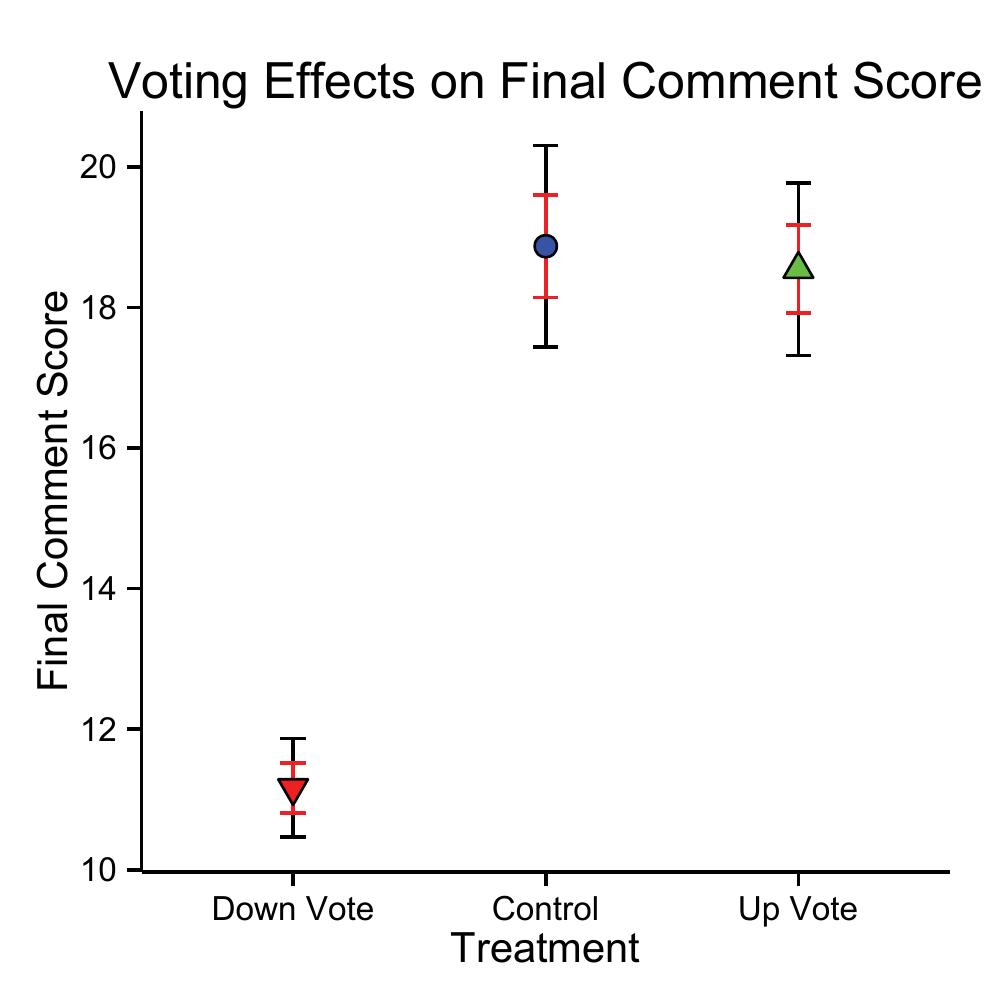}
        \subcaption{Comments}
    \end{minipage}
        
    \caption{Final scores for artificially, randomly up-treated posts, down-treated posts, and scores for untreated posts in the control group are shown. Red inner error bars show the standard error of the mean; black outer error bars show the 95\% confidence interval. Fig. (a) shows the scores in the heavily skewed full distribution for posts. Fig. (b) shows the scores in the heavily skewed full distribution for comments, with significant decreases for down-treated comments when compared to the control group. }
    \label{fig:main_res}
\end{figure}



\begin{figure}[t]  
    \centering
    \begin{minipage}[b]{0.35\textwidth}
        \includegraphics[width=\textwidth]{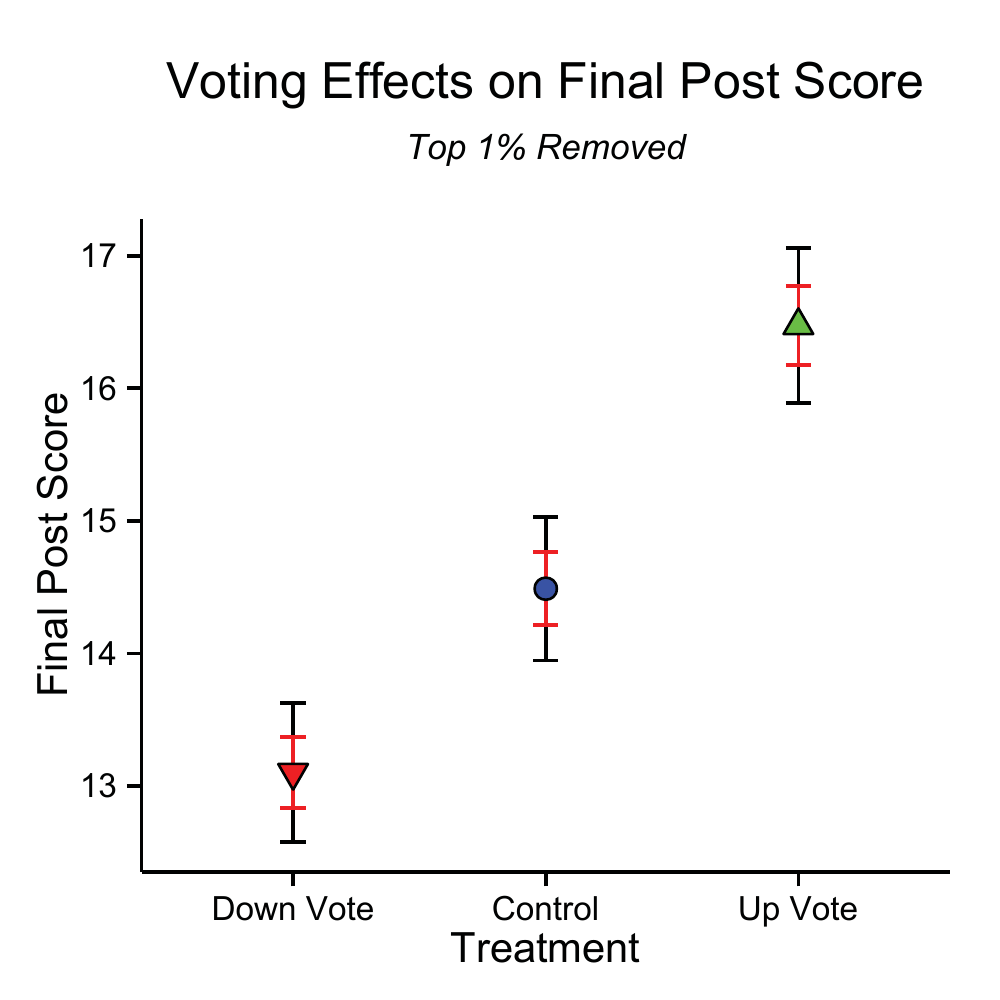}
        \subcaption{Posts}
    \end{minipage}
    ~
    \begin{minipage}[b]{0.35\textwidth}
        \includegraphics[width=\textwidth]{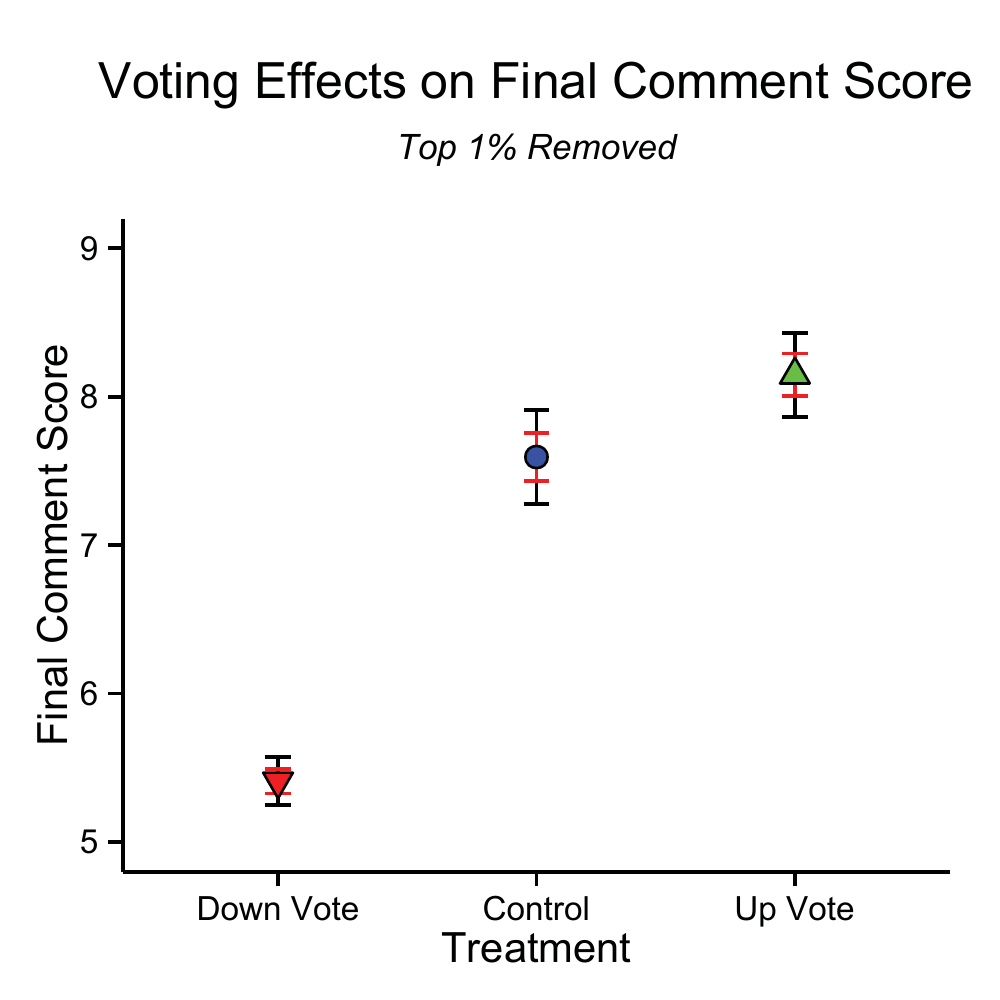}
        \subcaption{Comments}
    \end{minipage}
        
    \caption{Top 99\% of final scores for artificially, randomly up-treated posts, down-treated posts, and scores for untreated posts in the control group are shown. Red inner error bars show the standard error of the mean; black outer error bars show the 95\% confidence interval. When the highest 1\% of post scores are removed, the score distribution becomes much less skewed resulting in tighter error bounds, which further result in significant increases for up-treated posts and significant decreases for down-treated posts when compared to the control group. Again, when the highest 1\% of comment scores are removed, the score distribution becomes less skewed resulting in tighter error bounds, but with slight but not significant increases for up-treated comments when compared to the control group.}
    \label{fig:main_res_99}
\end{figure}

Figure \ref{fig:main_res} shows the full distribution of the final post scores and comment scores for each treatment group. Black outer error bars show the 95\% confidence interval and red inner error bars show the standard error of the mean. The full distribution of post scores in Figure~\ref{fig:main_res}(a) is extremely positively skewed with a skewness of $11.2$ and a kurtosis of $149.8$. If we remove the top 1\% highest scoring posts from the data set the skewness and kurtosis values drop to $6.5$ and $54.9$ respectively giving a better, although still skewed, view of the treatment effects. Figure~\ref{fig:main_res_99}(a) shows the distribution of the final post scores with the top 1\% of posts removed. In this case, the up-treated posts have a significantly higher final score, and the down-treated posts have a significantly lower final score.

The distribution of comment scores in Fig.~\ref{fig:main_res}(b) is even more positively skewed than the distribution of post scores with a skewness of $16.4$ and a kurtosis of $339.7$ but when the top 1\% highest scoring comments are removed, the skewness and kurtosis values dropped to $6.7$ and $58.1$, similar to the skewness and kurtosis for the distribution of post scores when the top 1\% highest scoring posts are removed. In this case, the down-treated comments have a significantly lower final score but the up-treated comments do not have a significantly higher final score.

Tests of statistical significance, {\em e.g.}, T-test, are known to improperly reject the null hypothesis when the data distribution is non-normal or highly skewed. This is indeed the case in our result set as is indicated by the abnormally high skewness and kurtosis scores. Removal of the top 1\% of scores is one way to unskew the data, hence the tightening of error bars and narrowing of confidence internals in Fig.~\ref{fig:main_res_99} as compared to the full results in Fig.~\ref{fig:main_res}.
Another way to unskew data is to take the log of each value in the distribution, which unfortunately removes negative scores from the analysis, a significant limitation for this line of work. 

Student's T-Test on the full set ({\em i.e.}, 100\%) of log-scores for posts also showed that the up-treated posts were significantly higher than the control group ($p = 1.69 \times 10^{-20}$), and that the down-treated posts were significantly lower than the control group ($p = 1.69 \times 10^{-09}$),  although scores less than or equal to 0 were removed to calculate the log of the final scores. For comments we find that Student's T-Test on log-scores demonstrated that up-treated posts were significantly higher than the control group ($p = 2.69 \times 10^{-24}$), and down-treated posts were significantly lower than the control group ($p = 9.18 \times 10^{-05}$). Unfortunately, the distribution of log-scores was still far from normal, so the T-Test is likely to give improper results.

With this in mind, we used the non-parametric, 1-dimensional Kolmogorov-Smirnov (K-S) test as well as the Mann-Whitney  $U$ (M-W) Test to determine the significance between treatments and control. Both the M-W and the K-S tests are nonparametric tests to compare two unpaired groups of data. They each compute p-values that test the null hypothesis that the two groups have the same distribution. They do have some important differences though. The M-W test operates by ranking all the values from low to high, and then computes a p-value that depends on the differences between the mean ranks of the two groups. The K-S test compares the cumulative distribution of the two data sets, and computes a p-value that depends on the largest difference between the two distributions. The differences between the two tests are important, but they both compute p-values that can be used to judge the statistical significance of the treatment effects. Thus, we will display the results of both tests.

The K-S Tests showed that the final score distribution of all up-treated posts were more positively skewed than posts in the control group (K-S test statistic: $0.08$; $p < 2.2 \times 10^{-16}$), which were more positively skewed than down-treated posts (K-S test statistic: $0.11$; $p < 2.2 \times 10^{-16}$). 
The same K-S test on comment scores shows significantly higher final scores for up-treated comments and down-treated comments ($p < 2.2 \times 10^{-16}$). The reason that the p-value of the K-S statistic is reported as being less than $2.2 \times 10^{-16}$ is because floating point underflow error prevents a more precise calculation in the R-based K-S test calculator.

Finally, we performed the independent 2-group M-W Test comparing treatments (up-treated and down-treated) with the control. We again find significant differences comparing the up-treated post scores to the control ($p=5.9 \times 10^{-53}$) and the down-treated post scores to the control ($p=7.8 \times 10^{-73}$). The same M-W Test on comments also showed significant differences in the final scores of up-treated comments compared to the control group ($p = 5.57 \times 10^{-15}$), and significantly different final scores in down-treated comments compared to the control group ($p = 7.52 \times 10^{-8}$).

\begin{figure}[t]
    \centering
    \begin{minipage}[b]{0.5\textwidth}
        \includegraphics[width=\textwidth]{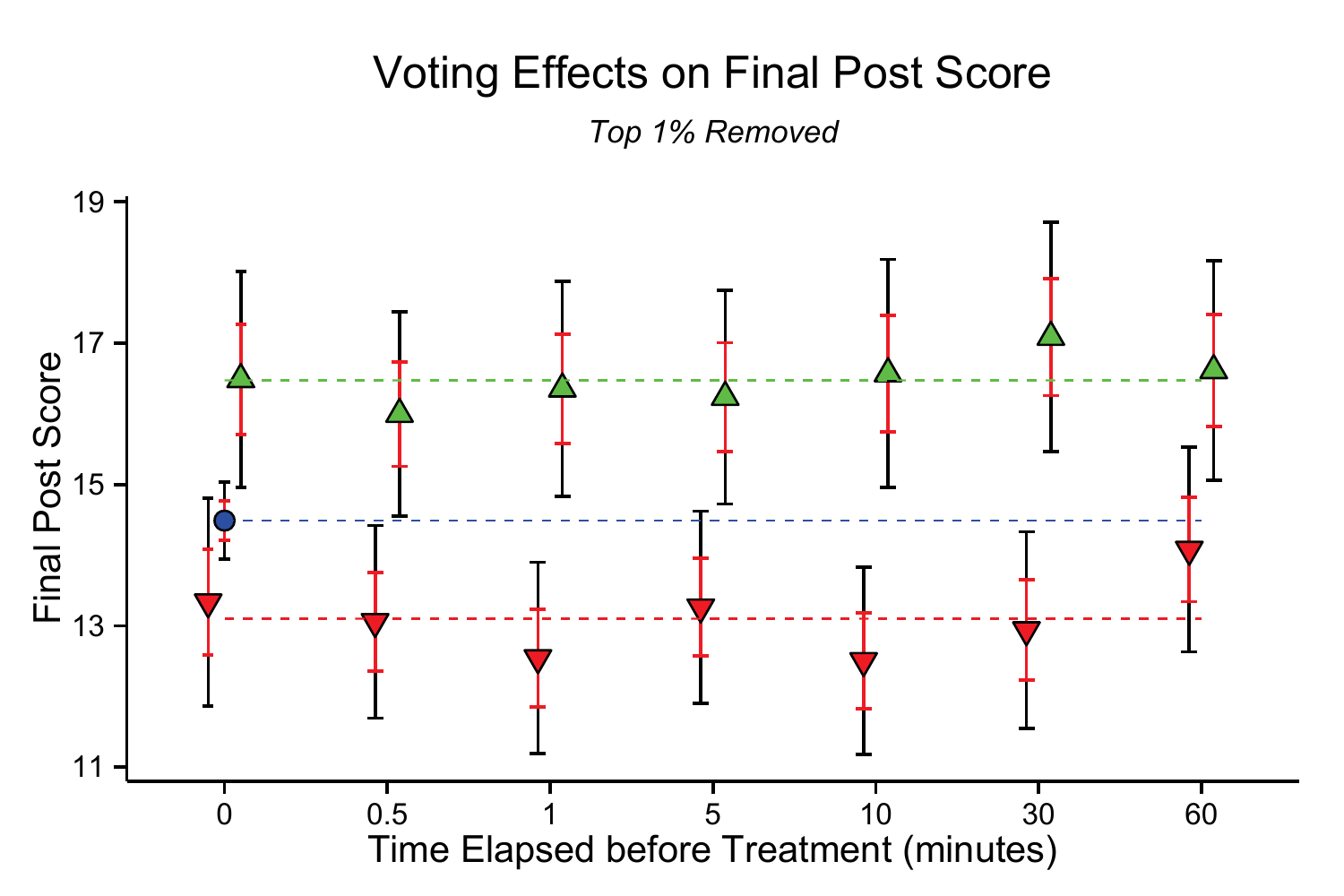}
        \subcaption{Posts}
    \end{minipage}%
    \begin{minipage}[b]{0.5\textwidth}
        \includegraphics[width=\textwidth]{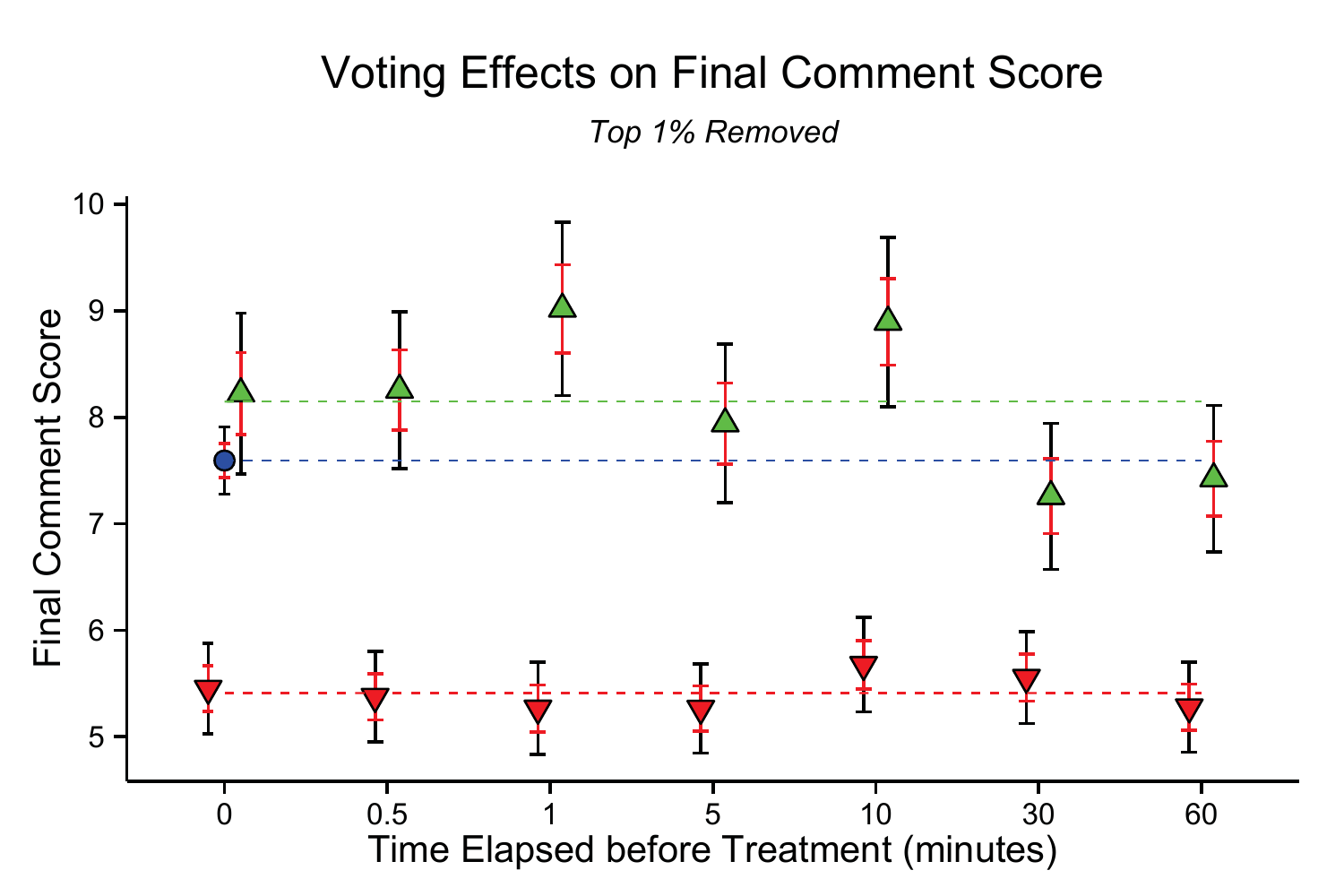}
        \subcaption{Comments}
    \end{minipage}
    
    \caption{Final scores separated into their respective treatment delay intervals. Fig. (a) shows final scores for artificially, randomly up-treated posts, down-treated posts, and scores for untreated posts in the control group and Fig. (b) shows final scores for artificially, randomly up-treated comments, down-treated comments, and scores for untreated comments in the control group. Horizontal lines show the overall mean of each treatment group. The top 1\% of scores were removed to un-skew the score distribution.}
    \label{fig:pctdec}
\end{figure}

In general, an up-vote increases a post's score on the site which increases its visibility according to the default ranking algorithms. The increased visibility of the post makes it more likely to be viewed by others. However, making a post more visible does not necessarily mean that it will receive more up-votes and continue to increase or even maintain its visibility; it may instead receive down-votes, thereby decreasing the posts visibility. That is, until we consider that the vast majority of votes cast on Reddit are up-votes and downvoting is actually discouraged unless the post is spam, off-topic, or otherwise improper. Thus, we are confident that the increase in the final post score after positive vote manipulation in the presence of popularity ranking mechanics is largely due to the increase in visibility due to the treatment up-vote.

Comments, in contrast, have a vastly different visibility mechanism than posts. Reddit comment threads are hierarchical, wherein the default ``best'' (highest up-vote to down-vote ratio) ordering mechanism sorts comments among its siblings only. The visibility of a comment in the hierarchy depends not only on its ordering among its siblings but also the rank of any parent or ancestor comments it has. Because our voting mechanism selected the most recent comment, it may be the case that the selected comment was a child or other descendant of a highly visible comment. As such, it may be the case that the treated comment was already highly visible by its relative position in the comment hierarchy. Unfortunately, we did not record the relative position of each treated comment and are unable to find correlation between relative visibility and treatment effects.

Another difference in the comments experiment is that, by default, only the top 200 comments are visible. By selecting ``rising'' posts, our collection methodology makes it highly likely that the comment that we select is within the first 200, and is therefore at least initially visible. Unfortunately, for large comment threads a single down-treatment may be enough to make the comment no longer visible under default orderings. This is probably why down-treated comments have such a low overall score compared to up-treated or control groups.

\subsection{Delay Effects}

Up-votes and down-votes for post receiving treatments were performed after a 0, 0.5, 1, 5, 10, 30 or 60 minute delay chosen at random, and Figures~\ref{fig:main_res} and~\ref{fig:main_res_99} does not distinguish between the effects of vote-treatments performed after the various delay periods. Figure~\ref{fig:pctdec} separates the results for posts from Figure~\ref{fig:main_res_99}(a) and Figure~\ref{fig:main_res_99}(b) into their respective treatment delay groups in Figure~\ref{fig:pctdec}(a) and Figure~\ref{fig:pctdec}(b), respectively. We expected that immediate votes would have a larger effect than votes performed after a long delay. However, these results show, surprisingly, that a delay in treatment generally did not have a significant effect on the mean outcome of a post's final score. 

\begin{table}[b]
\centering
\resizebox{0.99\linewidth}{!}{
\begin{tabular}{l c r|c c c c c c c}
                 &   &     & 0                            & 0.5 & 1  & 5 & 10 & 30 & 60 \\ \hline
\multirow{8}{*}{K-S} & \multirow{4}{*}{Post} & $\color{green} \blacktriangle$  & \specialcell{$D=0.087$\\$p<2.2\textsc{e}^{-16}$}  & \specialcell{$D=0.087$\\$p<2.2\textsc{e}^{-16}$}  & \specialcell{$D=0.087$\\$p<2.2\textsc{e}^{-16}$}  & \specialcell{$D=0.083$\\$p<2.2\textsc{e}^{-16}$}  & \specialcell{$D=0.082$\\$p<2.2\textsc{e}^{-16}$}  & \specialcell{$D=0.087$\\$p<2.2\textsc{e}^{-16}$}  & \specialcell{$D=0.078$\\$p<2.2\textsc{e}^{-16}$} \\

    &  &  $\color{red}\blacktriangledown$ & \specialcell{$D=0.119$\\$p<2.2\textsc{e}^{-16}$}  & \specialcell{$D=0.110$\\$p<2.2\textsc{e}^{-16}$}  & \specialcell{$D=0.110$\\$p<2.2\textsc{e}^{-16}$}  & \specialcell{$D=0.112$\\$p<2.2\textsc{e}^{-16}$}  & \specialcell{$D=0.119$\\$p<2.2\textsc{e}^{-16}$}  & \specialcell{$D=0.097$\\$p<2.2\textsc{e}^{-16}$}  & \specialcell{$D=0.099$\\$p<2.2\textsc{e}^{-16}$} \\ \cline{2-10}

    & \multirow{4}{*}{Com.} & $\color{green} \blacktriangle$  & \specialcell{$D=0.0.43$\\$p=1.8\textsc{e}^{-07}$}  & \specialcell{$D=0.063$\\$p=8.9\textsc{e}^{-16}$}  & \specialcell{$D=0.051$\\$p=3.2\textsc{e}^{-10}$}  & \specialcell{$D=0.045$\\$p=7.0\textsc{e}^{-08}$}  & \specialcell{$D=0.049$\\$p=1.6\textsc{e}^{-09}$}  & \specialcell{$D=0.038$\\$p=1.3\textsc{e}^{-05}$}  & \specialcell{$D=0.021$\\$p=0.04\dag$} \\

    &  &  $\color{red}\blacktriangledown$ & \specialcell{$D=0.0.49$\\$p=1.6\textsc{e}^{-08}$}  & \specialcell{$D=0.050$\\$p=9.7\textsc{e}^{-09}$}  & \specialcell{$D=0.049$\\$p=2.1\textsc{e}^{-08}$}  & \specialcell{$D=0.047$\\$p=3.5\textsc{e}^{-08}$}  & \specialcell{$D=0.038$\\$p=2.9\textsc{e}^{-05}$}  & \specialcell{$D=0.031$\\$p=0.1\textsc{e}^{-02}$}  & \specialcell{$D=0.037$\\$p=4.0\textsc{e}^{-05}$} \\ \hline

\multirow{4}{*}{M-W} & \multirow{2}{*}{Post} & $\color{green} \blacktriangle$  & $p=6.1\textsc{e}^{-14}$  & $p=1.4\textsc{e}^{-18}$  & $p=6.2\textsc{e}^{-18}$  & $p=1.8\textsc{e}^{-13}$  & $p=2.2\textsc{e}^{-11}$  & $p=6.2\textsc{e}^{-15}$  & $p=8.6\textsc{e}^{-12}$ \\

    &  &  $\color{red}\blacktriangledown$ & $p=3.3\textsc{e}^{-22}$  & $p=2.5\textsc{e}^{-17}$  & $p=4.5\textsc{e}^{-24}$  & $p=2.6\textsc{e}^{-21}$  & $p=9.9\textsc{e}^{-27}$  & $p=4.3\textsc{e}^{-15}$  & $p=1.5\textsc{e}^{-11}$ \\ \cline{2-10}

    & \multirow{2}{*}{Com.} & $\color{green} \blacktriangle$  & $p=1.1\textsc{e}^{-05}$  & $p=3.7\textsc{e}^{-10}$  & $p=1.5\textsc{e}^{-06}$  & $p=1.8\textsc{e}^{-05}$  & $p=7.3\textsc{e}^{-09}$  & $p=3.7\textsc{e}^{-04}$  & $p=0.11\dag$ \\

    &  &  $\color{red}\blacktriangledown$ &$p=0.1\textsc{e}^{-02}$  & $p=0.8\textsc{e}^{-02}$  & $p=0.3\textsc{e}^{-03}$  & $p=3.7\textsc{e}^{-05}$  & $p=0.17\dag$  & $p=0.31\dag$  & $p=0.1\textsc{e}^{-02}$ \\ \hline

\end{tabular}
}
\caption{Results of Kolmogorov-Smirnov (K-S) and Mann-Whitney $U$ (M-W) Tests on complete result set ({\em i.e.}, with top 1\% included). $\color{green} \blacktriangle$ represents tests comparing up-treatment with the control group; $\color{red} \blacktriangledown$ represents tests comparing down-treatment with the control group. $\dag$ indicates results that are not statistically significant at the 99\% confidence level.}
\label{tab:ci}
\end{table}

Unfortunately, displayed error bounds and confidence intervals, which are computed from Student's T-Test, have little meaning when the data is so highly skewed; K-S tests shown in Table~\ref{tab:ci} again showed that all up-treated posts were more positively skewed than posts in the control group and that the effects generally diminished as the delay interval increased. Similarly, the control group was more positively skewed than the down-treated posts, but the effects were mixed as the delay interval increased. 

As for comments, the K-S test results were more mixed in Table~\ref{tab:ci}, but still mostly statistically significant. The up-treated comments were significantly more positively skewed than the control group comments, and the down-treated comments resulted in a significantly lower score. Interestingly, the p-values of the comment scores diminished as the delay grows longer, meaning that the vote treatment on comments are not effective a half-hour or an hour after the comment has been made. In short, timely voting on a comment is more important than timely voting on a post on average.

M-W tests of statistical significance, also shown in Table~\ref{tab:ci}, demonstrate that post treatments have a significant effect across all delay periods, and that this effect only slightly diminishes (if at all) when the delay approaches 1 hour. 

As for comment treatments, the M-W tests showed significance results similar to those from the K-S tests. Namely, the effect of up-treatment, as measured by the p-value scores, diminished as the delay grew bigger and led to an insignificant effect when the delay was 1 hour. The effect of down-treatment was significant for short delay periods, but was not significant for delays of 10 minutes and 30 minutes, and was only barely significant for delays of 1 hour.

The results from the statistical tests on the comment treatments from Table~\ref{tab:ci} and Figure~\ref{fig:pctdec}(b) appear to be in conflict. Figure~\ref{fig:pctdec}(b) seems to show that negative treatments have a big effect on the final outcome of the comment for all delay levels, while positive-treatments have a little effect. However, proper statistical tests show that the truth is more nuanced. 

\begin{figure}[t]
    \centering
    \begin{minipage}[b]{\textwidth}
        \centering
        \includegraphics[width=0.19\textwidth]{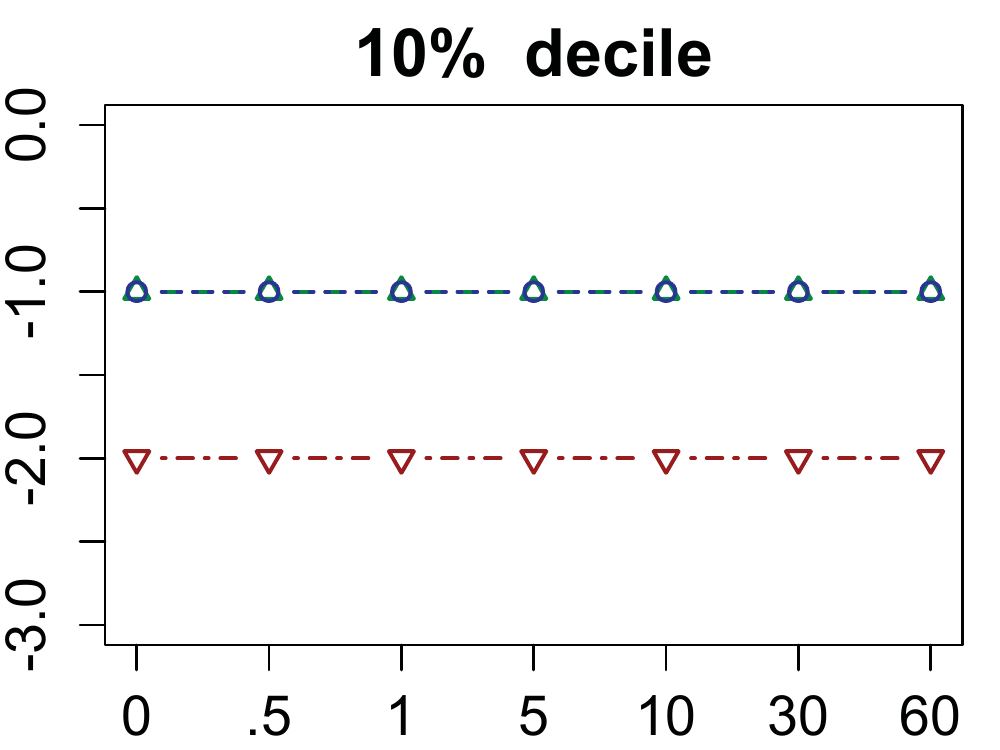}
        \includegraphics[width=0.19\textwidth]{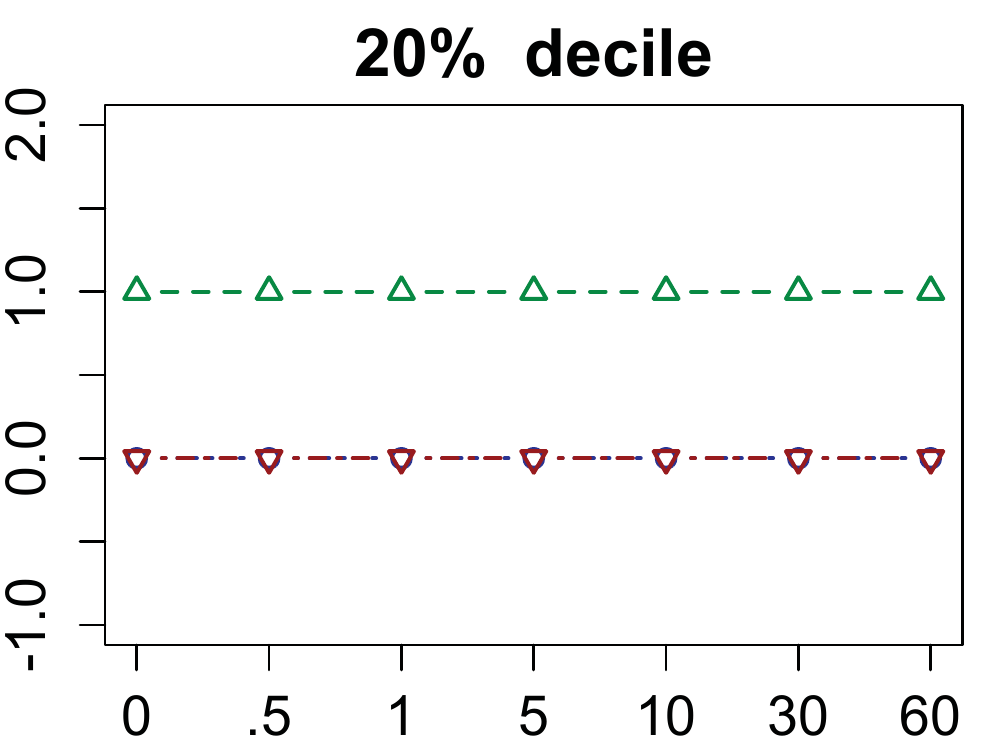}
        \includegraphics[width=0.19\textwidth]{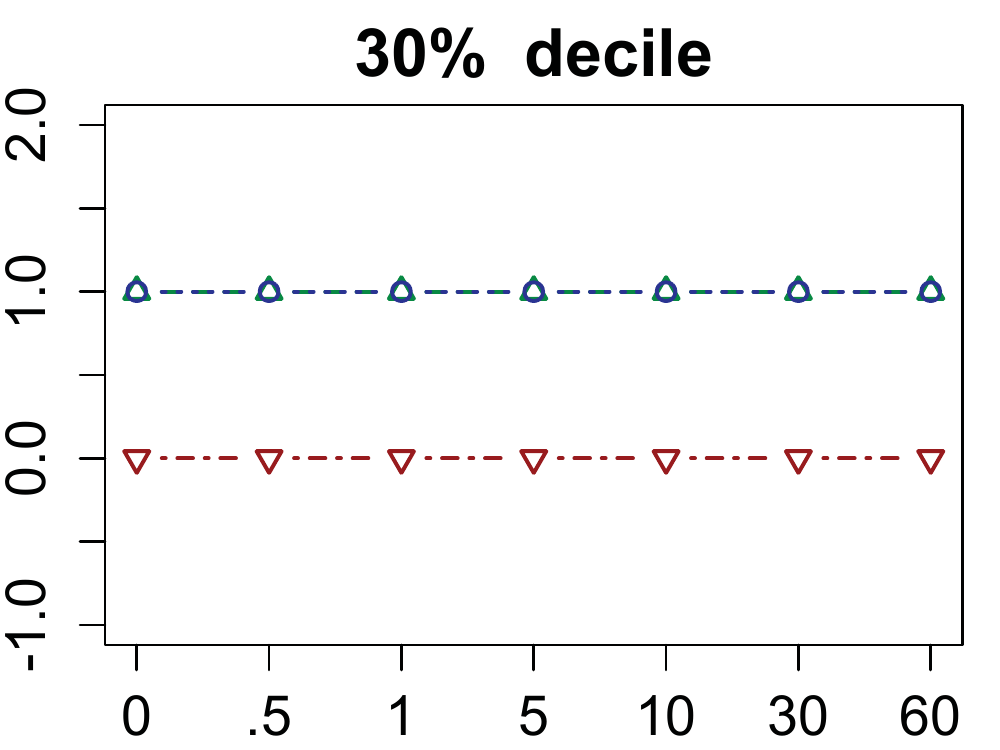}
        \includegraphics[width=0.19\textwidth]{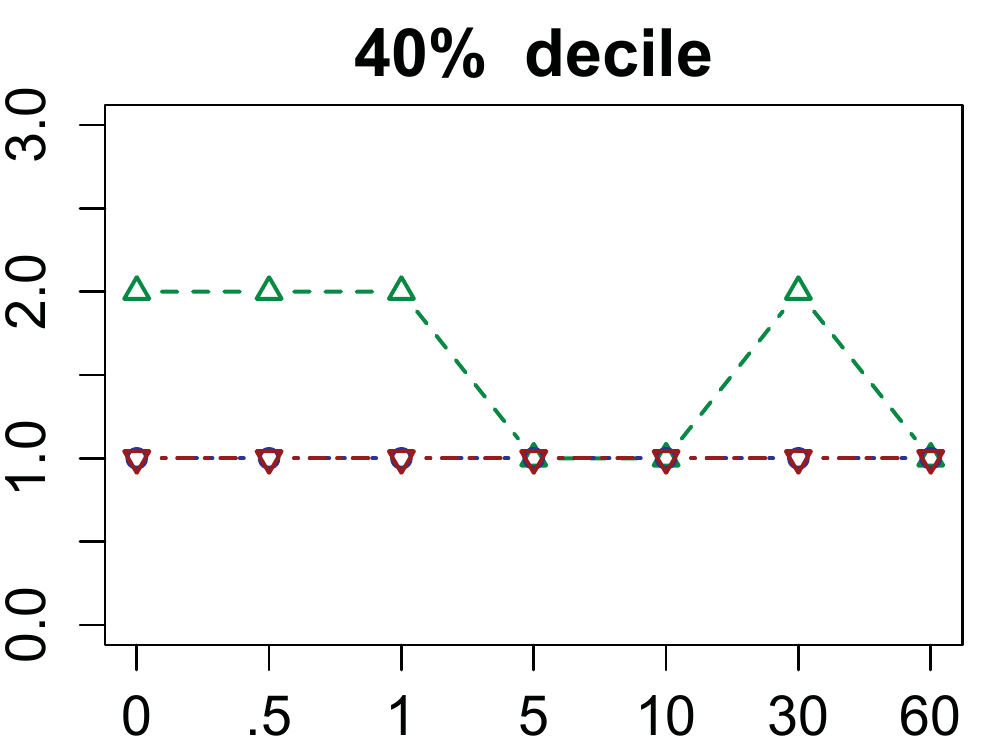}
        \includegraphics[width=0.19\textwidth]{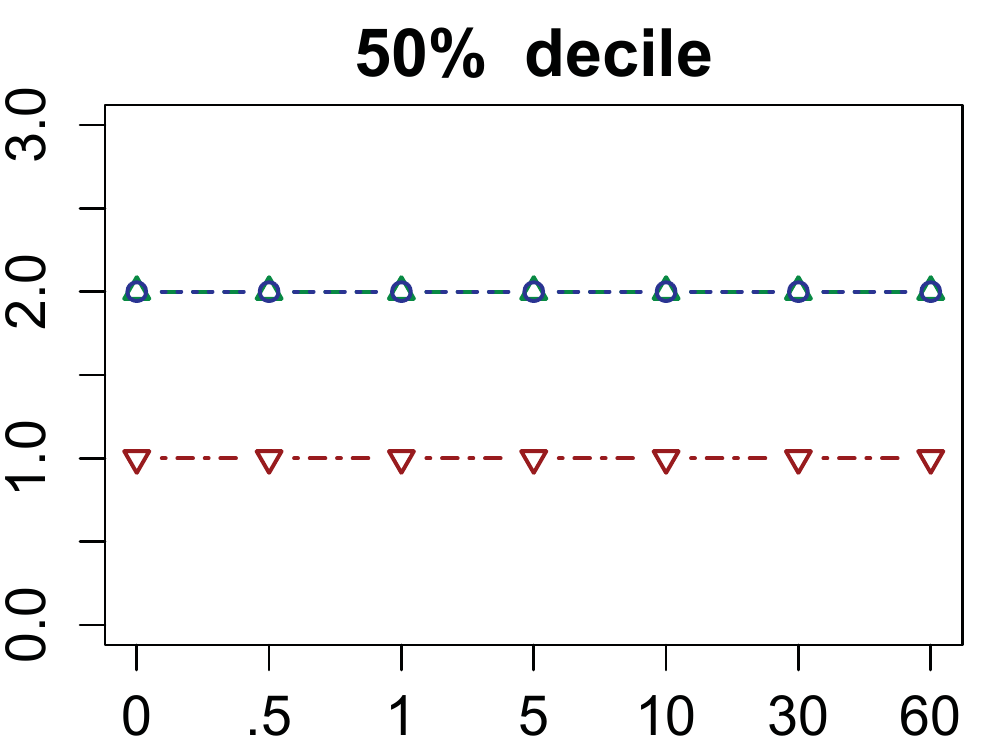}
        ~
        \includegraphics[width=0.22\textwidth]{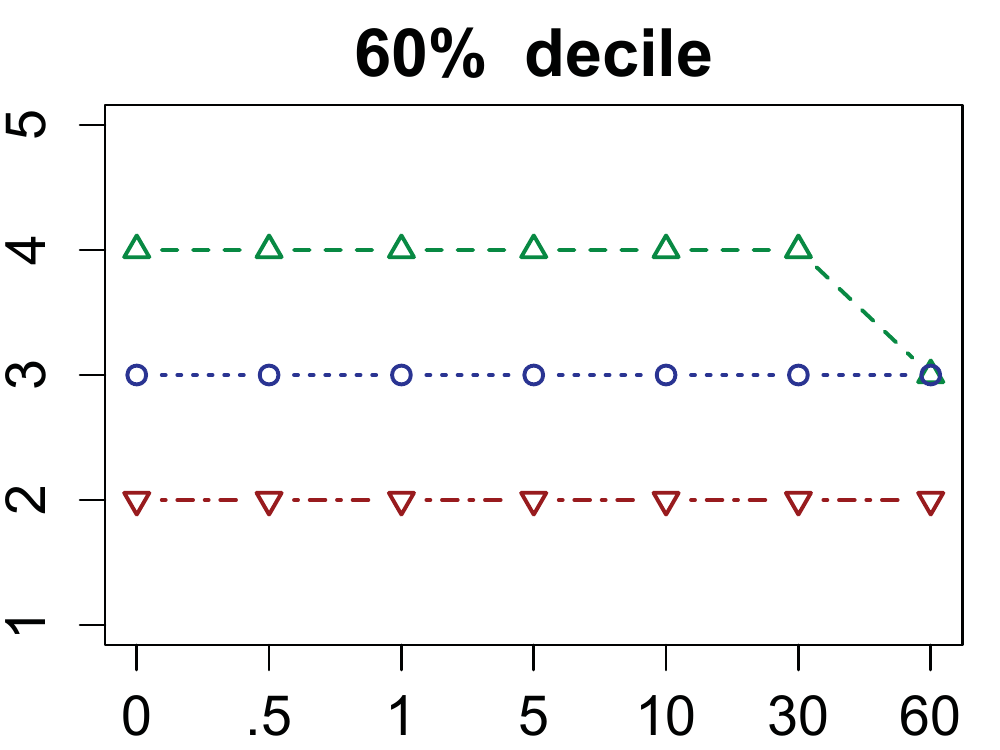}
        \includegraphics[width=0.22\textwidth]{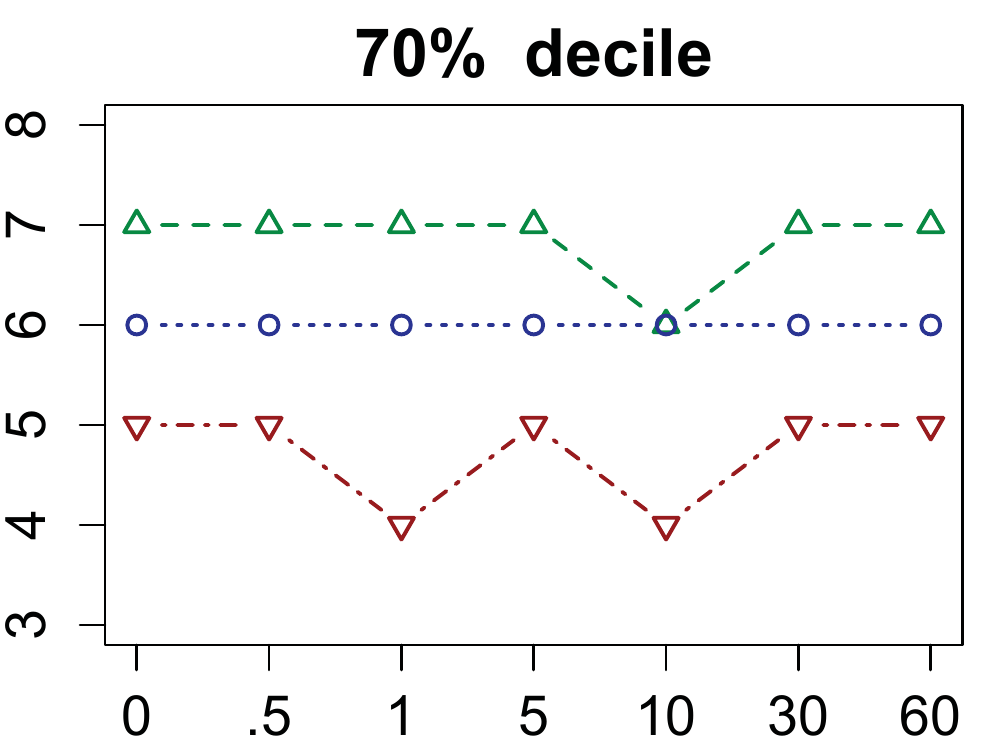}
        \includegraphics[width=0.22\textwidth]{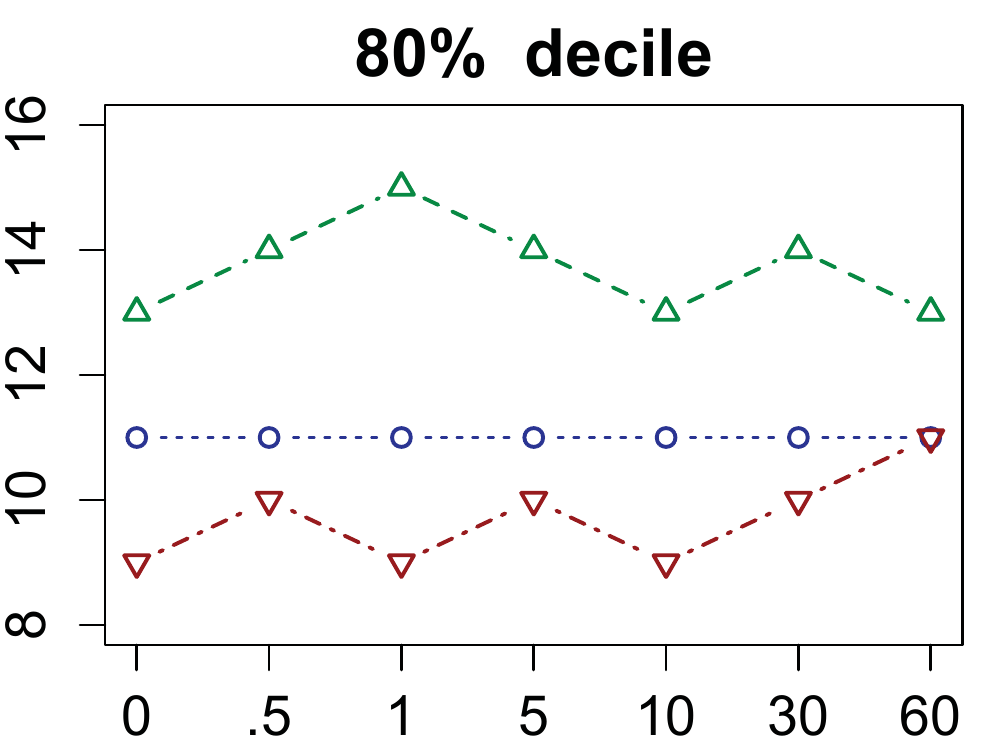}
        \includegraphics[width=0.22\textwidth]{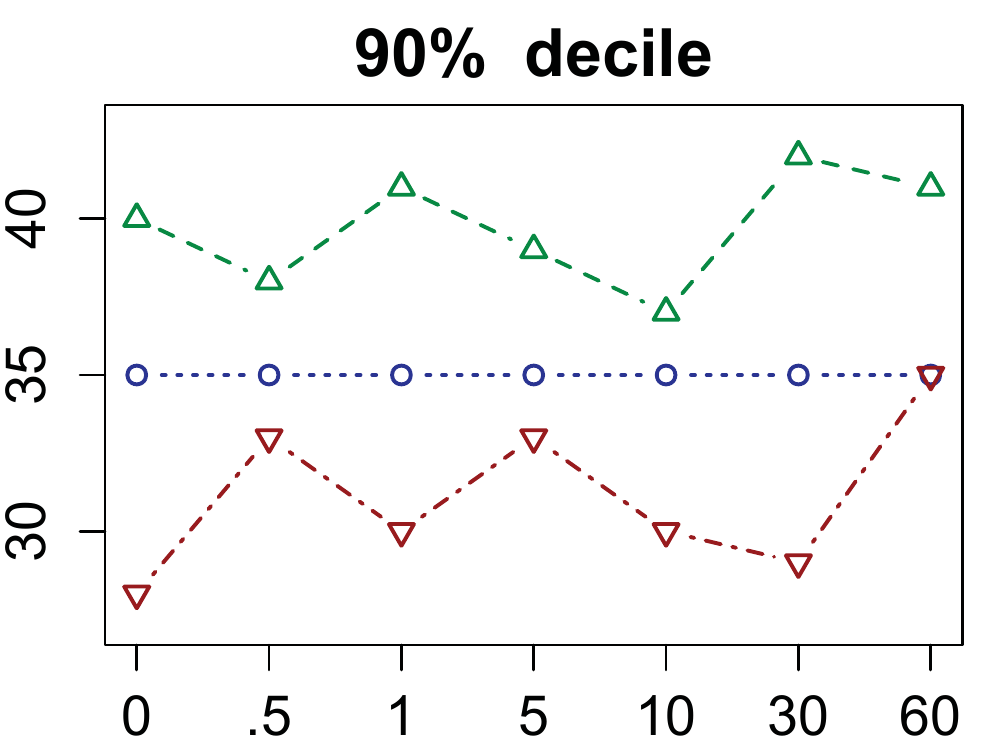}
        \subcaption{Posts}
    \end{minipage}
    \begin{minipage}[b]{\textwidth}
        \centering
        \includegraphics[width=0.19\textwidth]{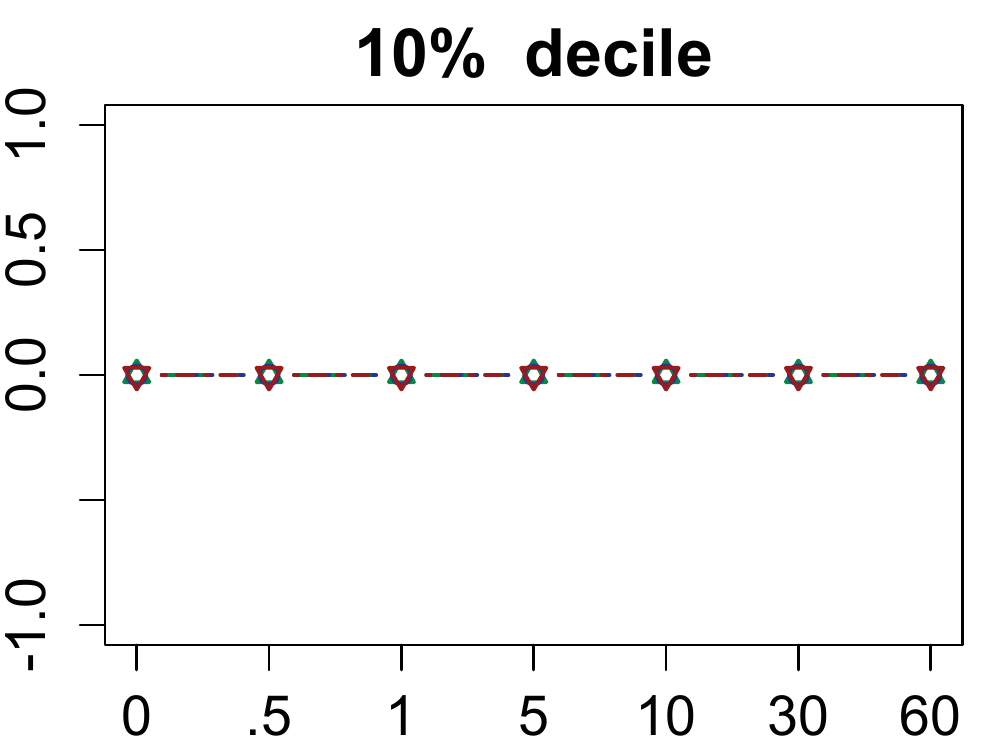}
        \includegraphics[width=0.19\textwidth]{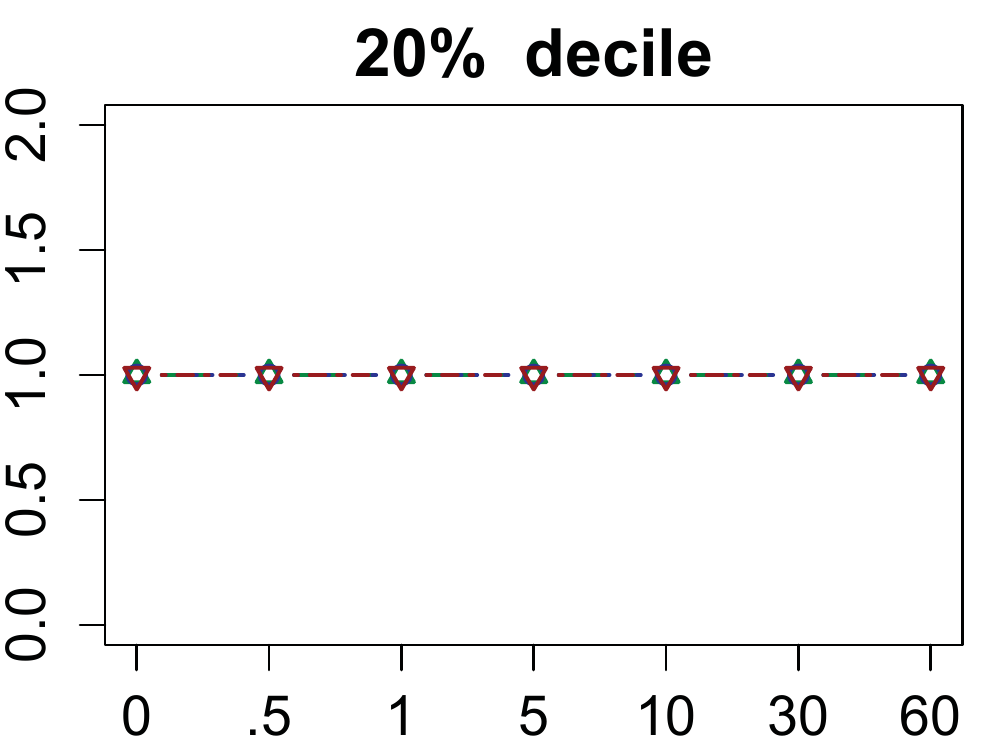}
        \includegraphics[width=0.19\textwidth]{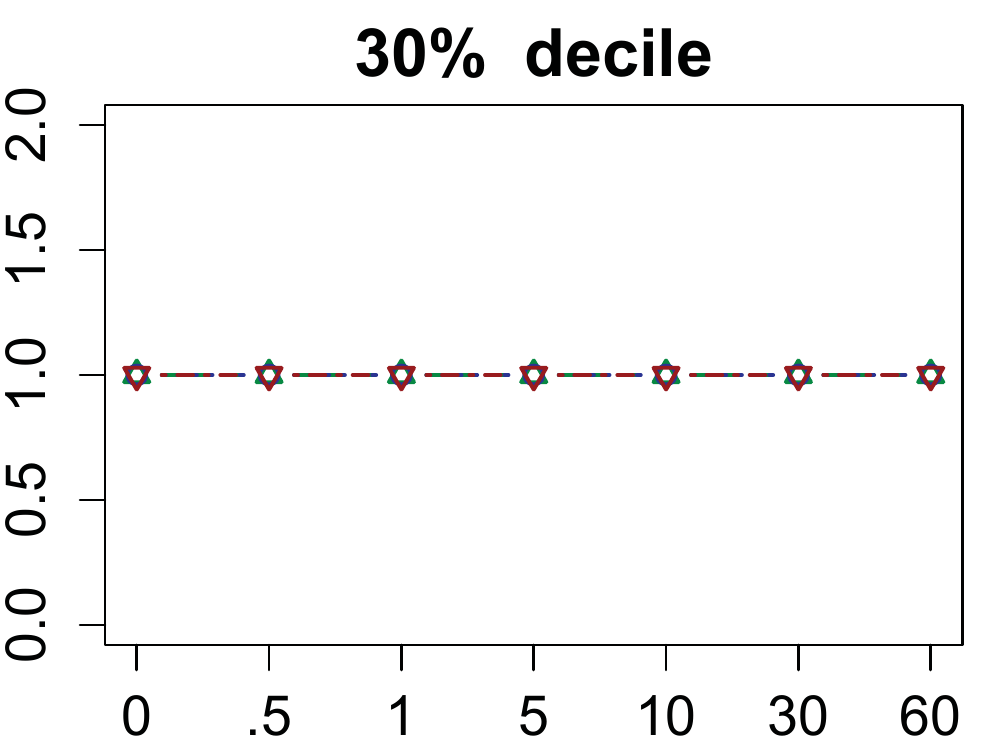}
        \includegraphics[width=0.19\textwidth]{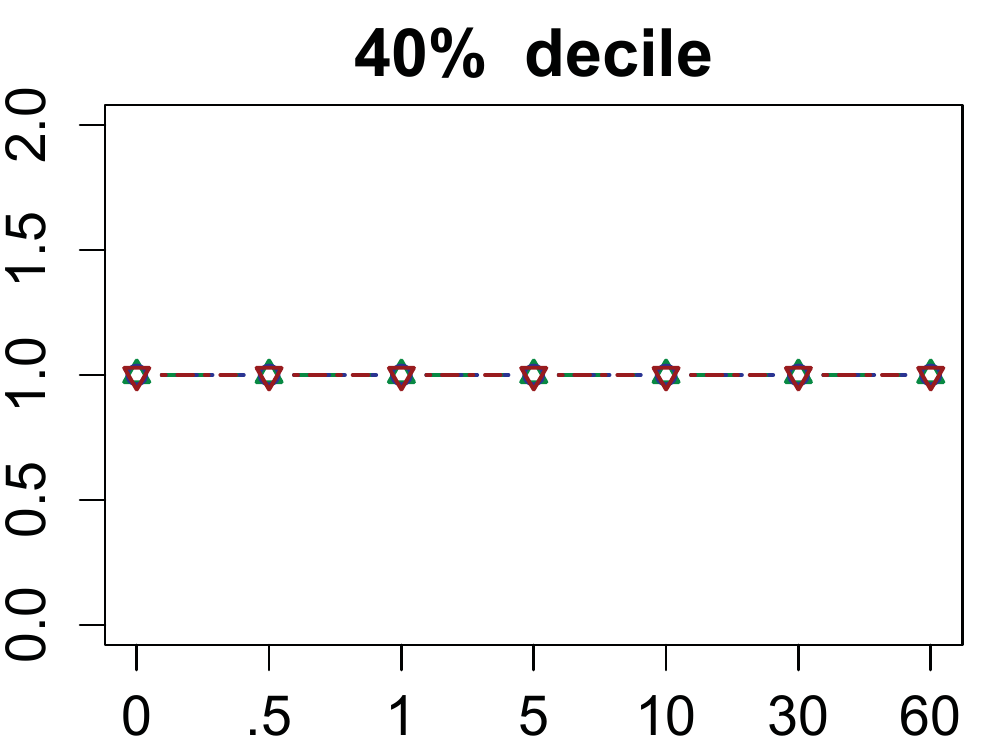}
        \includegraphics[width=0.19\textwidth]{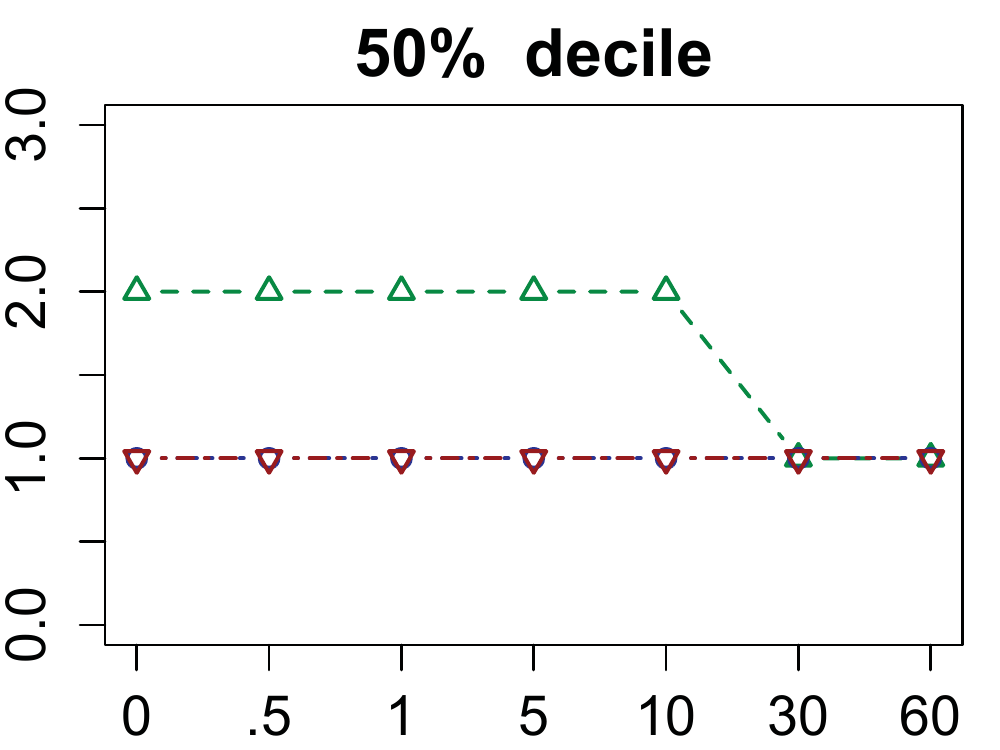}
        ~
        \includegraphics[width=0.22\textwidth]{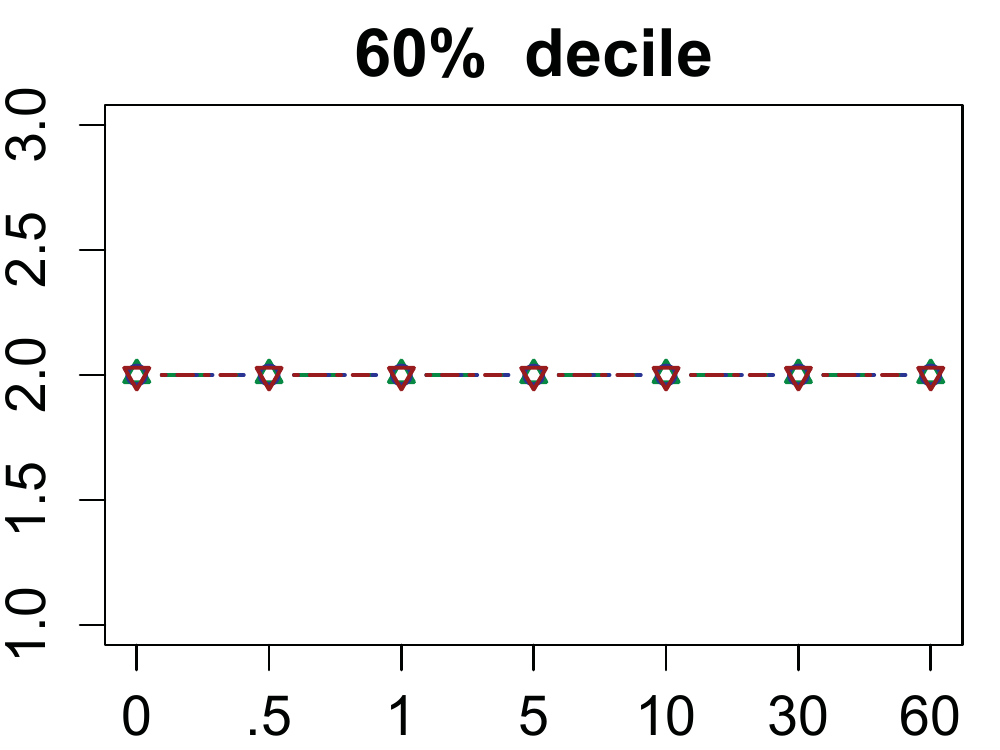}
        \includegraphics[width=0.22\textwidth]{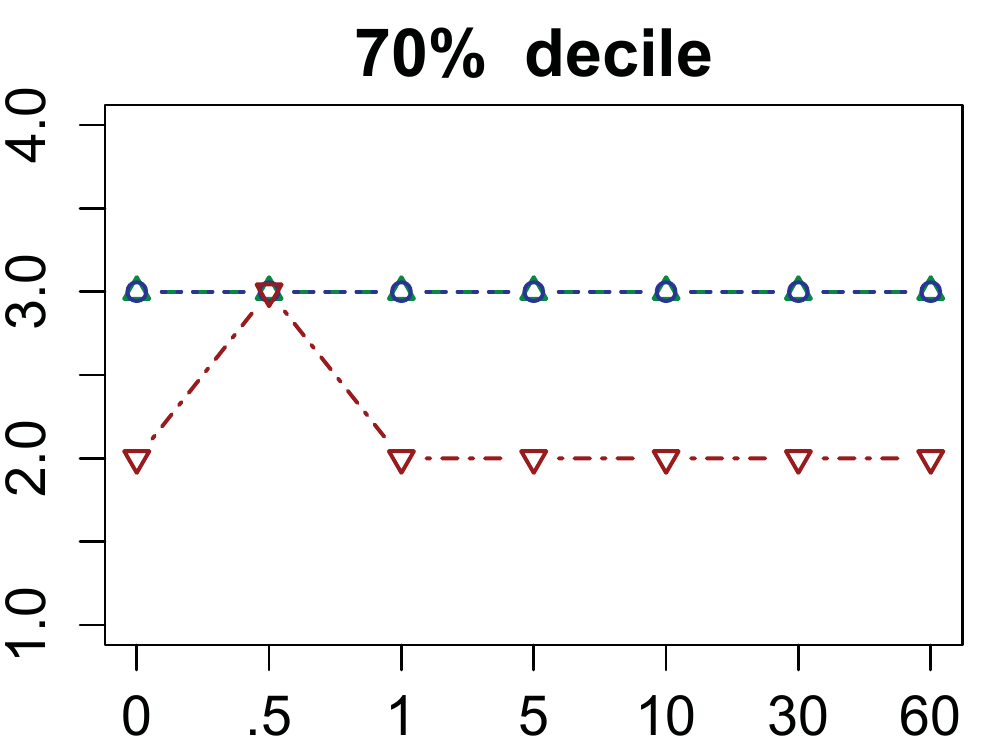}
        \includegraphics[width=0.22\textwidth]{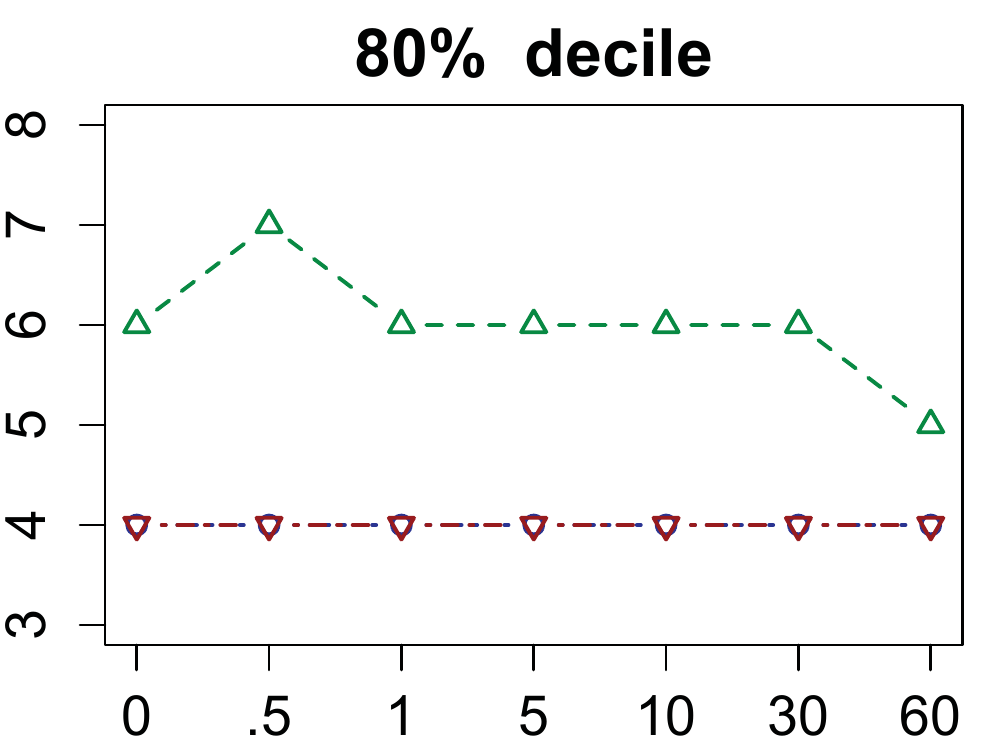}
        \includegraphics[width=0.22\textwidth]{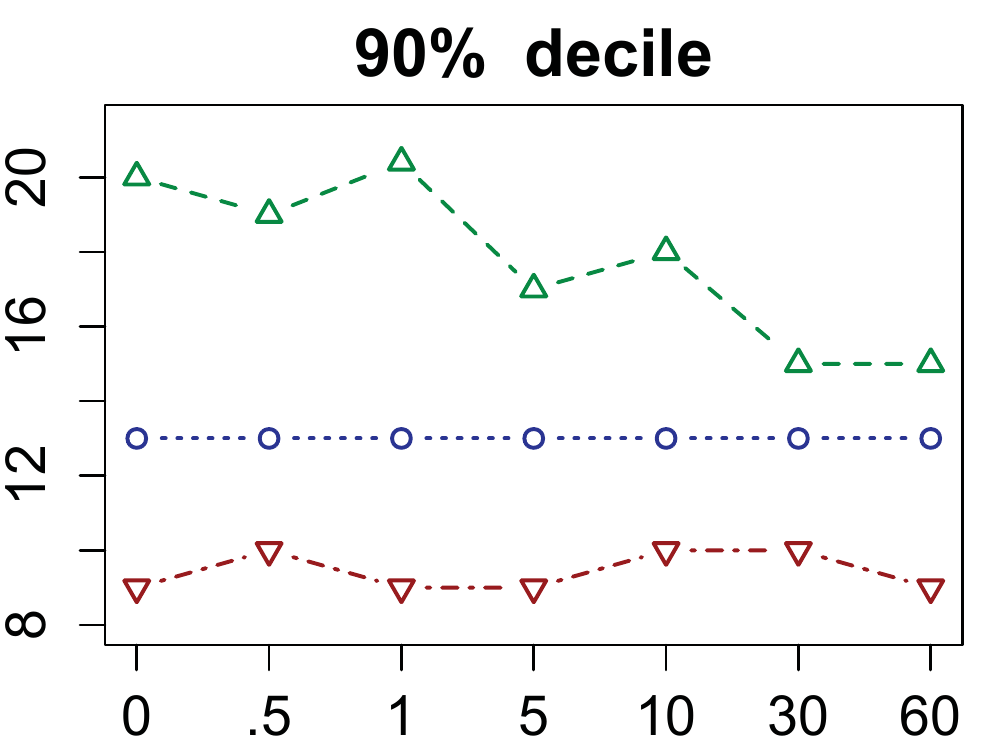}
        \subcaption{Comments}
    \end{minipage}
    \caption{The middle 9 deciles of final scores for each treatment according to their delay intervals. These results show that most posts and comments receive a median score of 2 or less, and that treatment has the most effect in the higher deciles of the score distribution.}
    \label{fig:deciles}
\end{figure}

Ultimately, with this type of data, the best way to show aggregate results is through n-tile plots. With this in mind, Figure~\ref{fig:deciles} shows the inner-deciles of the results as a function of their treatment delay. Taken together these results show graphically what the tests of statistical significance imply: that up-treated posts tend to score more highly than the control group, and that down-treated posts tend not to not score as highly as the control group. The decile plots also show that the majority of posts (deciles $\le$ 50\%) receive at most a final score of 2, and that most comments never receive any votes at all. 

\subsection{Reaching the Front Page}

\begin{figure}[t]
    \centering
    \begin{minipage}[b]{0.5\textwidth}
        \includegraphics[width=\textwidth]{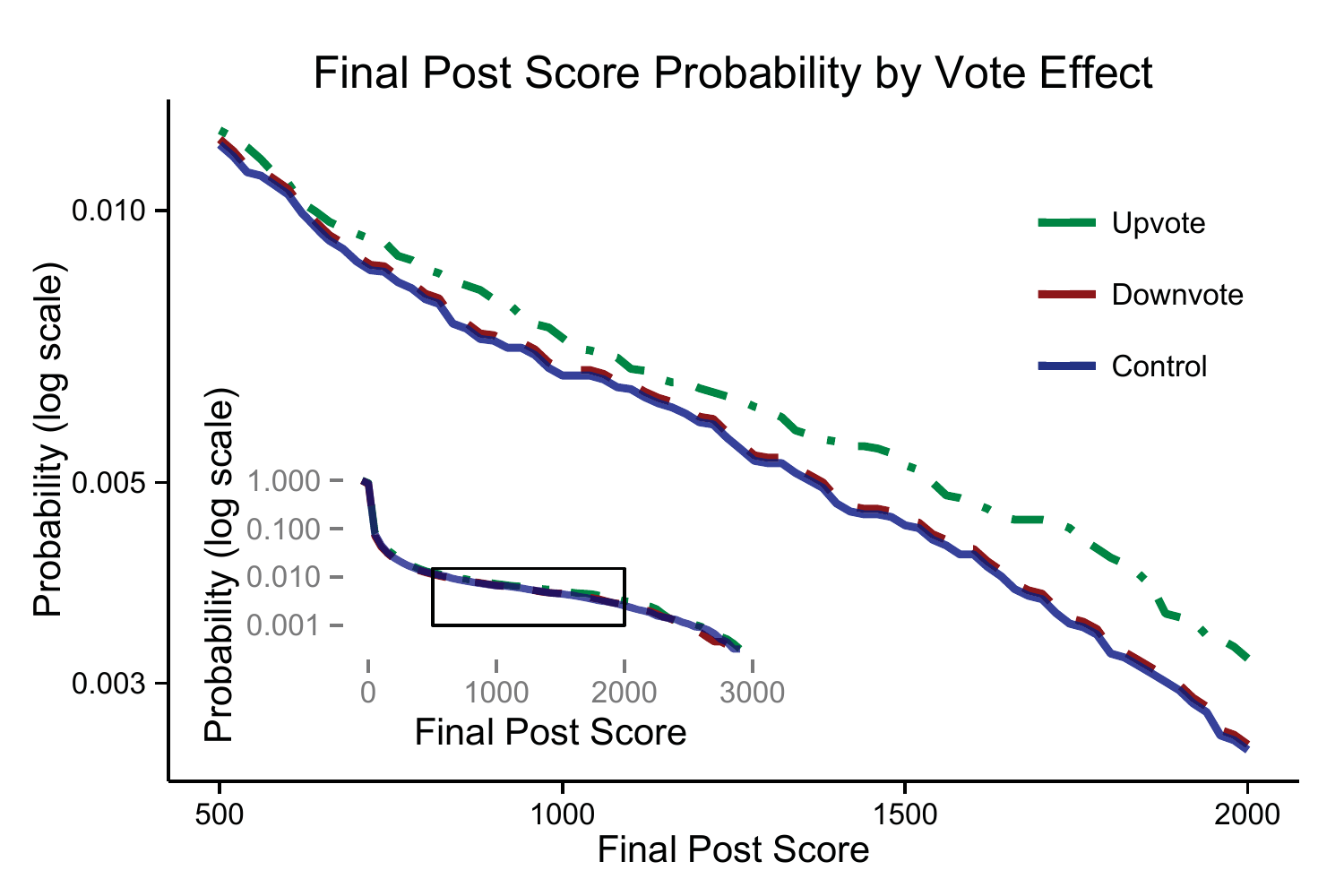}
        \subcaption{Posts}
    \end{minipage}%
    \begin{minipage}[b]{0.5\textwidth}
        \includegraphics[width=\textwidth]{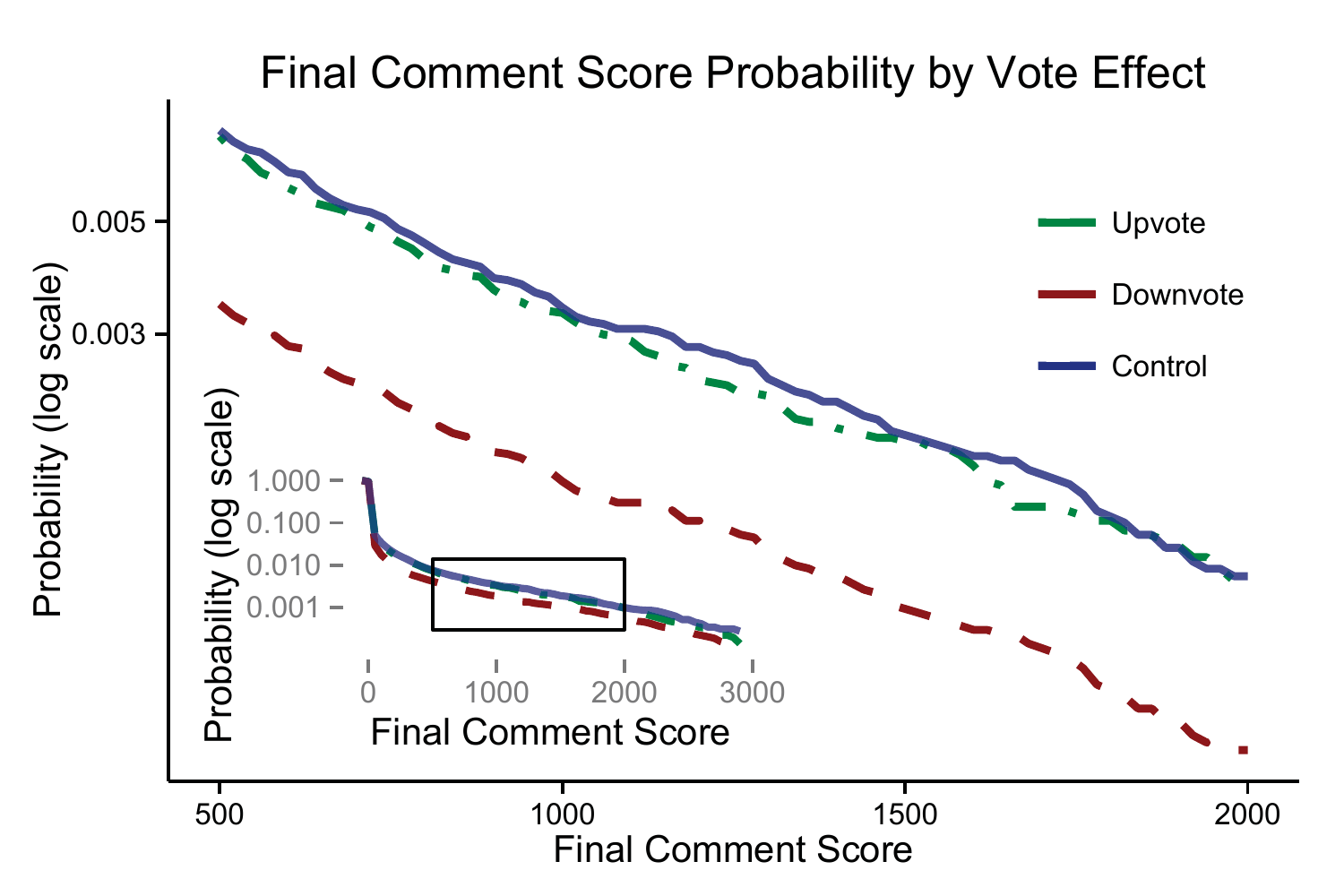}
        \subcaption{Comments}
    \end{minipage}
    
    \caption{The probability of a post (a) or a comment (b) receiving a corresponding score by treatment type. The inset graph shows the complete probability distribution function. The outer graph shows the probability of a post receiving scores between 500 and 2000 -- an approximation for {\em trending} or {\em frontpage} posts. Up-treated posts are 24\% more likely to reach a score of 2000 than the control group.}
    \label{fig:final_score_prob}
\end{figure}

Overall, the results suggest that an up-treatment increases the probability that a post will result in a high score relative to the control group, and that down-treatments decrease that probability relative to the control group. However, on Reddit and other social news sites only a handful of posts become extremely popular. On Twitter and Facebook this is generally referred to as a {\em trending} topic, but on Reddit the most popular posts are the ones that reach the front page. Unfortunately, reaching the front page is a difficult thing to discern because each user's homepage is different, based on the topical subreddits to which the user has subscribed. Nevertheless, we crudely define a post as having become popular, {\em i.e.}, is trending, on the frontpage, etc., if it has a score of more than 500. Using this definition, Figure~\ref{fig:final_score_prob}(a) shows the probability that post reaches a given final score under the two treatment conditions. These probability distribution functions are monotonically decreasing, positively skewed, and show that up-treatment results in a large departure from the control group for posts and down-treatment results in a large departure from the control group for comments. However, despite our earlier claims of up-treatment and down-treatment symmetry on post results, these results show that, in the upper limits of the distribution, down-treatments do not effect the final score results. These results mean that, compared to the control group, an up-treated post is 7.9\% more likely to have a final score of at least 1000, and an up-treated post is 24.6\% more likely to have a final score of at least 2000.

The probability that a comment reaches a high score is generally lower than the probability of a post reaching the same high score because posts are generally more viewed and voted on than comments. Indeed, in order to even view the comments, a user must first view, or at least click-on, the post. Also, lower rated comments or comments with multiple levels of ancestor comments above them are often hidden until a user chooses to reveal them. Figure~\ref{fig:final_score_prob}(b) shows the probability that a comment reaches a given final score under the two treatment options as in Figure~\ref{fig:final_score_prob}(a). Interestingly, we find that an up-treatment has very little effect on the probability of a comment reaching a high score; yet, a down-treatment has a dramatic negative effect on that probability.

\subsection{Subreddit Effects}

\begin{figure}[t]
    \centering
    \begin{minipage}[b]{0.5\textwidth}
        \includegraphics[width=\textwidth]{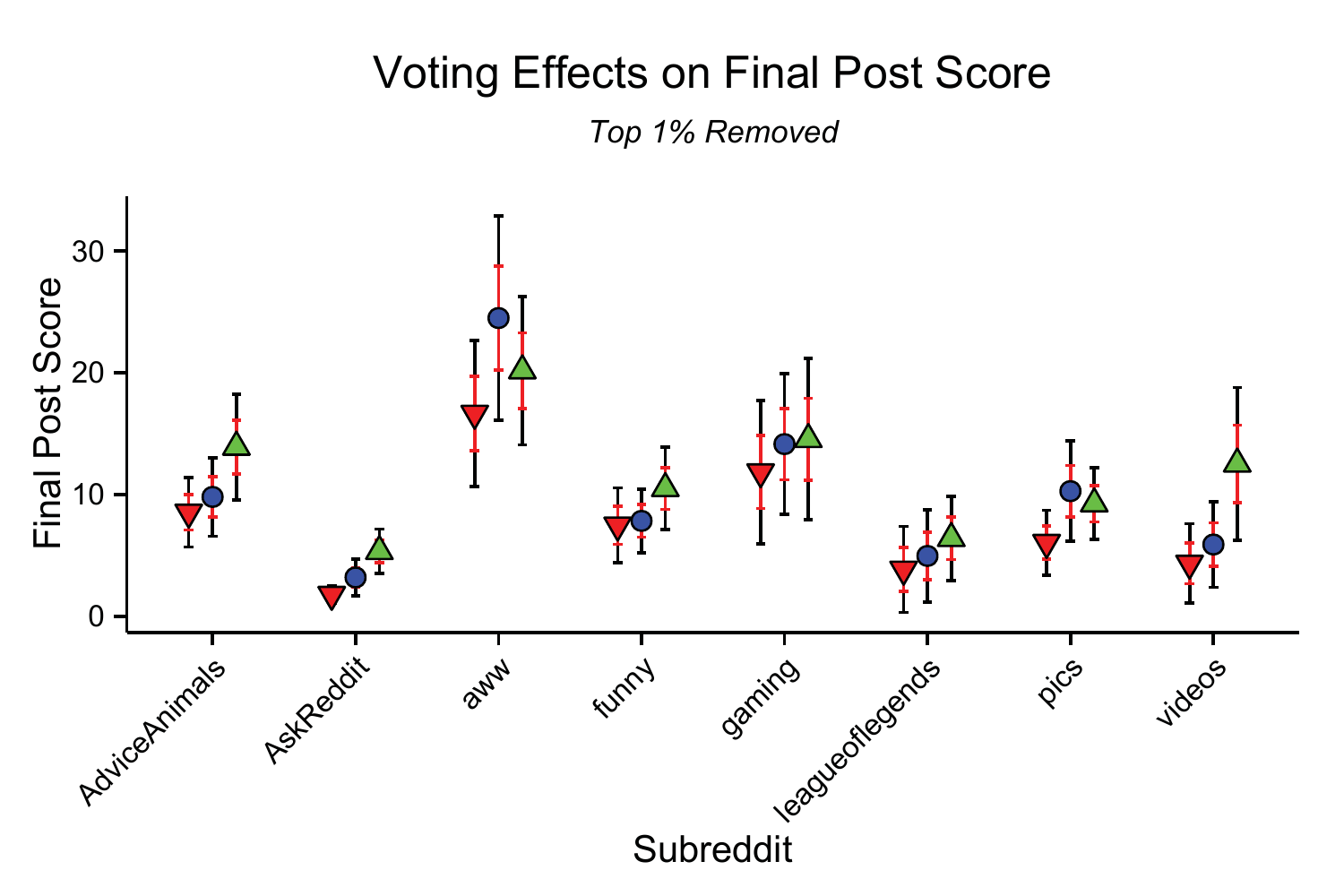}
        \subcaption{Posts}
    \end{minipage}%
    \begin{minipage}[b]{0.5\textwidth}
        \includegraphics[width=\textwidth]{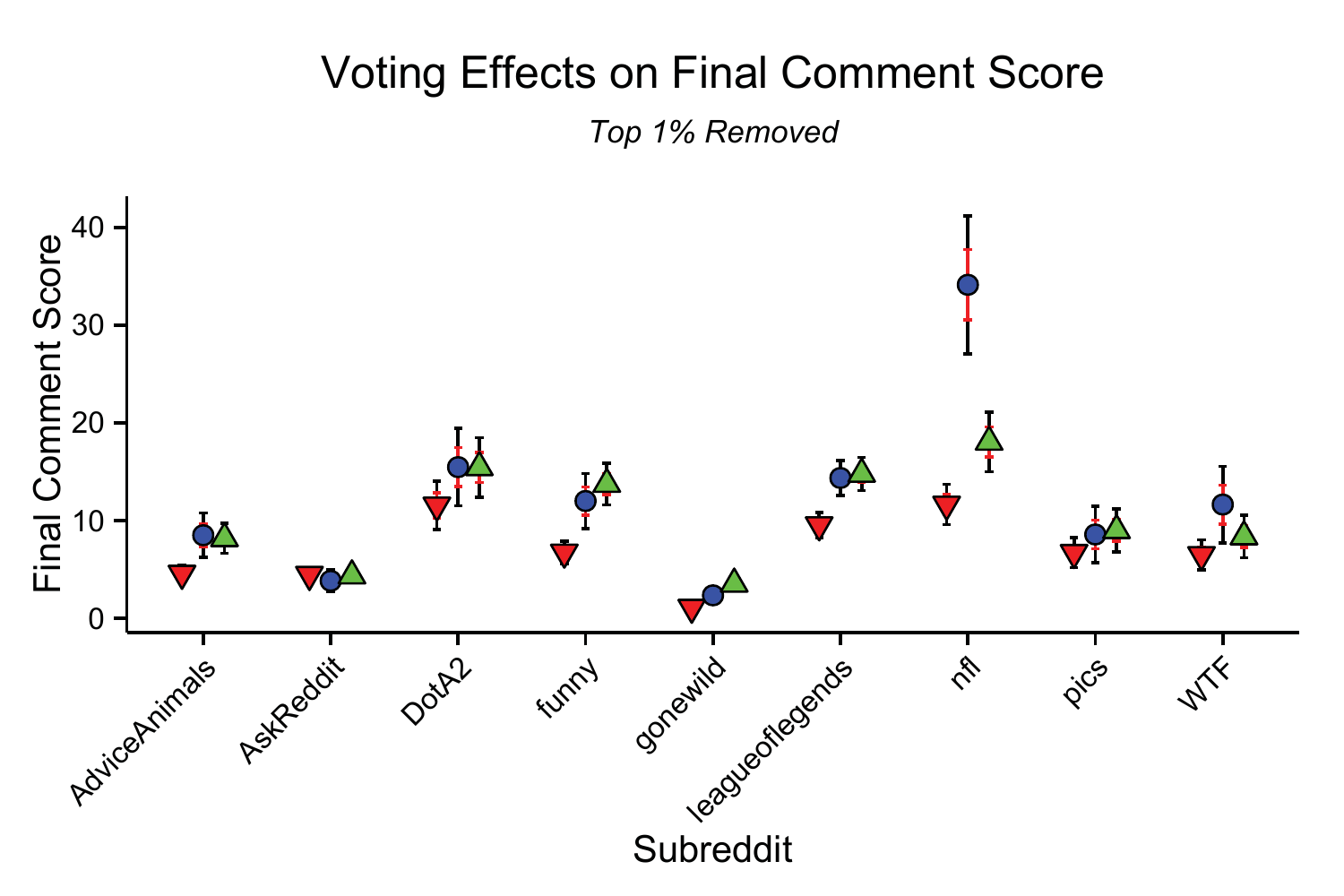}
        \subcaption{Comments}
    \end{minipage}
    
    \caption{Mean scores of down-treated, control group and up-treated posts in the top 8 most active subreddits in Fig. a and the  mean scores of down-treated, control group and up-treated comments in the top 9 subreddits from which we collected the most data in Fig. b, {\em i.e.}, subreddits that are (a) most active, (b) most often appearing as ``rising'' on Reddit. Black outer error bars show the 95\% confidence interval and red inner error bars show error.}
    \label{fig:mean_by_SR}
\end{figure}

We finally investigated treatment effects in the top 10 most frequent subreddits. These do not necessarily correspond to the top 10 most popular subreddits or the subreddits with the most comments. Rather, they are the subreddits to which posts are most frequently submitted or whose posts are most frequently ranked first on Reddit's ``rising'' ranking system due to our data collection methodology.

From the top 10 subreddits for posts, we removed \textsf{politic} and \textsf{friendsafari} and from the top 10 subreddits for comments, we removed \textsf{friendsafari}. These subreddits were removed from our analysis because posts in \textsf{politic} are automatically submitted by a computer program, and because posts and comments in \textsf{friendsafari} cannot be down-voted according to the subreddit rules. Thus, only 8 subreddits for posts and 9 subreddits for comments are shown in Figure~\ref{fig:mean_by_SR} which illustrates the effects of treatment on post and comment scores on average within top 10 subreddits. 

Figure~\ref{fig:mean_by_SR}(a) and Mann-Whittney test results show significant positive effects on post scores in \textsf{AdviceAnimals}, \textsf{AskReddit} and \textsf{videos}, and significant negative effects on post scores in \textsf{AskReddit} and \textsf{pics}. These results illustrate similar symmetric effects that we found on posts overall. Voting effects on comment scores within subreddits are shown in Figure~\ref{fig:mean_by_SR}(b). While we find that down-treatments typically result in significantly lower final comments scores compared to the control, up-treatments rarely result in significantly higher final comment scores as shown by Figure~\ref{fig:mean_by_SR}(b). 

Within the top 500 subreddits for posts, we found that 22\% had significant up-treatment effects, 21.6\% had significant down-treatments, and 5.4\% of subreddits had significant differences in both the up-treatment and down-treatment results when compared to the control group. There was also no correlation in the top 500 between up-treatment significance and number of submissions the subreddit received ($r^2 = 0.014$; p-value = $0.007$), nor down-treatment significance and number of submissions the subreddit received ($r^2 = 0.010$; p-value = $0.026$).

\section{Related Work}
Although this is the first in-vivo Reddit experiment, our work is motivated and informed by multiple overlapping streams of literature and build on substantial prior work from multiple fields such as: herding behavior from theoretical and empirical viewpoints~\cite{Salganik2006,Weninger2013}; social influence~\cite{Bakshy2009}; collective intelligence~\cite{Hirshleifer1995,Anderson2012a}; and online rating systems~\cite{Lu2008}.
A recent study by Muchnik \textit{et al} on a small social news Web site, similar to Reddit, found that a single up-vote/like on an online comment significantly increased the final vote count of the treated comment; interestingly, the same experiment also found that a single negative rating had little effect on the final vote count~\cite{Muchnik2013}.

In a separate line of work, Sorensen used mistaken omissions of books from the \textit{NY Times} bestsellers list to identify the boost in sales that accompany the perceived popularity of a book's appearance on the list~\cite{Sorensen2007}. Similarly, when the download counters for different software labels were randomly increased, Hanson and Putler found that users are significantly more likely to download software that had the largest counter increase~\cite{Hanson1996}. Salganik and Watts performed a study to determine the extent to which perception of quality becomes a ``self-fulfilling prophecy.'' In their experiment they inverted the true popularity of songs in an online music marketplace, and found that the perceived-but-false-popularity became real over time~\cite{Salganik2008}.

These experiments aim to determine the causal effect of social influence on rating behavior, as well as the mechanisms driving socio-digital influence. Although these experiments are first-of-a-kind, they are motivated and informed by multiple overlapping streams of literature and build on substantial prior work from multiple fields such as: herding behavior from theoretical~\cite{Bikhchandani1992,Banerjee1992,Hamilton1971} and empirical viewpoints~\cite{Salganik2006,Wu2008,Leskovec2007b,Celen2004,Anderson1997}; social influence in networks~\cite{Bakshy2009,Leskovec2006,Manski1993,Aral2009,Myers2012}; collective intelligence~\cite{Woolley2010,Brabham2008,Bikhchandani1992,Hirshleifer1995}; and online rating systems~\cite{Chevalier2006,Wu2008,Lu2008,Chen2008,Myers2012,Luca2011,Duan2008a,Duan2008,Hu2006,Li2008,Zhu2010a,Dellarocas2007}. Interestingly, most of the previous work is geared towards marketing science because of the close relationship between business and consumer opinion. 

The dynamics of online reviews, ratings and votes have received a lot of recent attention in the computing and marketing literature because the dynamics of online reviews for books, restaurants, hotels, etc. have become a vital business interest~\cite{Luca2011,Duan2008a,Duan2008,Hu2006,Li2008,Zhu2010a,Dellarocas2007}. Recent work in text mining is able to automatically determine the positivity and negativity of user-opinion~\cite{Lu2008,Lu2010,Lu2009} even among different aspects of a certain product (\textit{e.g.}, large can be a good thing when talking about portion size, but bad when talking about camera size)~\cite{Lu2011}. These papers attempt to codify ratings from plain, user-generated text and then determine relationships between the ratings and popularity.

Nonetheless, studies that aim to demonstrate the ease of online manipulation of ratings or voting tend to be limited. The biggest limitation is that these studies assume that the manipulators have full knowledge of the voting preferences of every user -- a valid assumption in theoretical work, but a meaningless assumption in real-world applications~\cite{Satterthwaite1975,Gibbard1973,Bartholdi1991,Bartholdi1989}. There is some work that considers manipulators who have a limited~\cite{Conitzer2011} or probabilistic~\cite{Ballester2009,Majumdar2004} knowledge of the voting preferences, but these assumptions are still too limiting for our purposes.

On the practical side, one obvious case of online manipulation is spam, particularly a new type of spam called \textbf{social spam}. Social spam is on the rise, with Nexgate Research reporting a tripling of social spam activity every six months~\cite{Nguyen2013} and BusinessWeek magazine reporting that only 40\% of online social media users are real people~\cite{Kharif2012}. There has been some practical work on detecting social spam in online social \textit{networking} Web sites~\cite{Tynan2012} like Facebook and Twitter, but not in social news platforms like Reddit and HackerNews. The largest and perhaps most effective type of social spam relies on social \textit{networks} to broadcast and propagate the advertisement or message~\cite{Zhang2012,Baeza-Yates2012,Ghosh2012,Markines2009}. These social network spammers are also the easiest to detect and shut down.  However, social news platforms are purposefully not social networks.

\section{Discussion}

In general, we find that the positive treatment of a single, random ``up-vote'' on a Reddit post has a corresponding positive herding effect that increases post scores on average and in the top limits of the heavily skewed score distribution but that a single, random ``up-vote'' on a Reddit comment had no significant positive herding effects. We further found that the negative treatment of a single, random ``down-vote'' on a post or comment has a corresponding negative herding effect that significantly decreased the post or comment scores on average, in contrast to the asymmetric findings of Muchnik \textit{et al.}~\cite{Muchnik2013}, who found no significant effects of a negative treatment. However, our results begin to resemble asymmetry in the top limits of the post score distribution meaning that a negative treatment does not decrease the probability that a post will receive a high score in the way that it does for comments. 

Separating treatments by their delay intervals did not yield a significant difference in effect overall. K-S and M-W tests found that up- and down-treatments for most delay intervals had significant effects compared to the control.. In general, the time that a vote is placed did not change the overall effect for post scores, but longer delays did diminish the effects that votes had on comment scores.

\subsection{Voting and Viewing Mechanics}
Research in social news manipulation has been shown a great deal of interest in recent years because of its centrality in shaping the news and opinion of society. There are several conflicting reports that now need to be teased apart. The work by Muchnik {\em et al.} showed positive herding effects but not negative herding effects~\cite{Muchnik2013} on non-Reddit social media comments ranked by recency, rather than popularity, and in the presence of a friendship social network. The voting and visibility mechanics of Reddit, which govern the data collection in this paper, are vastly different then the small or contrived experiments studied in earlier work. 

The post experiment and results are actually more in line with past research on ranking and visibility bias~\cite{Lerman2014} because of how the ranking mechanics of posts impact visibility. The results of our analysis as described above and the behavior of voting on Reddit, with an overwhelming majority of votes being upvotes and the discouragement of down-voting posts that are appropriate, support an increase in popularity from an increase in visibility. Thus, we are confident that \textbf{the increase in the final post score and the probability of reaching a high post score after positive vote manipulation in the presence of popularity ranking mechanics is largely due to the increase in visibility due to the treatment up-vote}.

\subsection{Vote-based Manipulation}

Collectively, this work, in the presence of other work on this subject~\cite{Lerman2014,Muchnik2013,Stoddard2015}, shows that votes determine visibility which, in turn, drives more votes. The 1\% rule, or its variants like the 90-9-1 rule, the 80/20 rule or the pareto principle, when applied to social media indicates that about 90\% of users only view content, 9\% of users edit content (including voting), and 1\% of users actively contribute new content~\cite{vanMierlo2014,Hargittai2008}. On all manner of vote-based social media platforms, the 10\% of users who actually vote are the ones who determine the kind of content that becomes widely visible and circulated among the remaining 90\% of the viewing public. Therefore, that active 10\% determines the ideas and opinions that the public is exposed to and influenced by.

Clearly, there is a huge incentive for opinion-pushers to manipulate the visibility of certain ideas and opinions on social media Web sites. There are several types of vote-based manipulation techniques that exist. We discuss a few of them here.

\paragraph{Vote Brigading}Vote brigading is when a large group of people all conspire to up-vote or down-vote a particular post or idea. This is not unique to Reddit, as Twitter has its Retweet armies that attempt to manipulate the velocity of some discussion in order to artificially force a topic to become a ``trending'' topic. The social media Web site Digg was particularly susceptible to vote brigading, wherein only users with many friends could ever hope to have a post reach the frontpage because the poster's friends would initially vote on the post in order to raise its visibility enough so that the wider community to see it.

Fortunately, most forms of this type of vote manipulation can be easily detected and stopped with spam detection and prevention techniques~\cite{Fayazi2015,Lee2014}. As part of a larger strategy, Reddit now encourages hyperlinks between subreddits to be tagged with a no-participation URL, which restricts access for non-subscribers of the subreddit to read-only, in order to prevent ``cross-subreddit contamination\footnote{\protect\url{http://www.reddit.com/r/NoParticipation}}.'' For example, a no-participation link from \textsf{/r/yankees} to a post in \textsf{/r/redsox} (a historical baseball rivalry) would prevent Yankees fans from downvoting posts that favor Red Sox fans.

\paragraph{Vote Nudging}Vote nudging is the type of vote manipulation that is studied in this paper and is the easiest, and most common, type of vote manipulation on social media. A post or comment is most susceptible to being ignored when it is young. Vote nudging is when someone asks a few friends to up-vote the post or comment in order to give it a positive boost during its initial appearance. After the initial boost, the post is left to grow normally.

As we have shown in this study, vote nudging can be extremely successful because the default ranking system gives higher visibility to posts with more, timely votes. Vote nudging also prevents instances when an unrelated user down-votes, and effectively kills a posts' changes of becoming visible, because a post with three or four up-votes may be able to withstand the effects of a down-vote better than a post with no up-votes.

It is difficult to say how much vote nudging happens in social media. It is common for users to have multiple accounts for this reason, but multiple votes from the same IP address is easy for spam prevention systems to catch. 

\paragraph{Reverse Vote Nudging}Reverse vote nudging is when a user down-votes all of the posts or comments that are similar to their post or comment in order to make a relative gain on competing content.  For example, if a user contributes a post about the winner of a baseball game to \textsf{/r/yankees}, several other users may also have contributed posts about the same baseball game at about the same time to \textsf{/r/yankees}. In order the increase the relative ranking of their own post, the user may down-vote all of the other posts by the other users thereby increasing the relative ranking of their submission.

Similarly, a user may wish to down-vote all of the posts or comments that are ranked just above the user's submission in order to increase the relative ranking of the user's submission. Using the same baseball example as earlier, if there are posts dealing with other Yankees content that are ranked just above the user's post, then the user may increase the ranking of their own post, and therefore increase its visibility, by downvoting the other content.

\subsection{Conformity and Influence}
Comment threads on Reddit are a unique supplement to the posted content. In fact, it is widely thought that most social media users, across all types of platforms, read the title of the post and skip directly to the comments section -- although this has not been empirically researched. Also, Reddit, Youtube, Twitter and many other social media platforms, to some extent, show the current score of each comment in the comment thread. Thus, the opinions and ideas expressed in each comment are given an explicit rating from the voting user base that is often viewed as the prevailing opinion of the overall population.

The Asch conformity experiments in the 1950's and onward showed that perceptions of popular opinion can have profound effects on individual perceptions of the truth~\cite{Asch1955}. Social comment threads frequently have instances where the highest scored comments represent an incorrect fact or are contrary to be the prevailing public opinion, perhaps due to comment manipulation discussed above. However, it is sometimes uncomfortable for many comment readers to hold opinions contrary to what they perceive to the the prevailing opinion. This disillusionment sometimes leads to a position change, but can also lead to a retreat inwards due to confirmation bias, which, in the worst case, leads to radicalization.

\subsection{Voting}

The nature of the manner in which social platforms rank items for viewing typically utilizes the ratings, in this case the post or comment scores, of the items being ranked. The results of our experiments show that random vote perturbations through vote treatments impact the scores of posts and comments on Reddit. These results underscore the need for counter measures against vote chaining and social engineering strategies as multiple artificial votes are likely to increase the herding effect. 

Finally, we bring attention back to what Eric Gilbert calls,  the `widespread underprovision of votes' in social media like Reddit~\cite{Gilbert2013}. Although our data does not draw these figures explicitly, we estimate that a very small number of the daily visitors to social media Web sites actually vote on the items they view. This seems to be an even further skewed anecdote of the 1-9-90 rule of social networking~\cite{vanMierlo2014}, and may be an underestimated reason behind the results presented in this paper.

\section{Acknowledgments}
We thank Michael Creehan for his help and discussion. This research is sponsored by the Air Force Office of Scientific Research FA9550-15-1-0003. The research was approved by University of Notre Dame institution review board and the Air Force Surgeon General's Research Compliance Office. Raw data files, and statistical analysis scripts are available on the corresponding authors Web site at \url{http://www3.nd.edu/~tweninge/data/reddit_report.html}. Reddit Inc was not involved in the experimental design, implementation or data analysis.

\end{document}